\documentclass[usenatbib]{mn2e}
\usepackage{graphicx,subfigure}

\title[Galaxy Haloes in Clusters]
{Simulations of the Effects of Stripping and Accretion on Galaxy Haloes in 
Clusters.}
\author[David M.\ Acreman, Ian R.\ Stevens, Trevor J.\ Ponman, Irini Sakelliou]
{David M.\ Acreman, Ian R.\ Stevens, Trevor J.\ Ponman, Irini Sakelliou \\
School of Physics and Astronomy, University of Birmingham,
Edgbaston, Birmingham, B15 2TT, UK\\
Email: da@star.sr.bham.ac.uk, irs@star.sr.bham.ac.uk,
tjp@star.sr.bham.ac.uk, irini@star.sr.bham.ac.uk}
\date{
 Accepted ..............................; 
 Received ..............................;
 in original form ..............................}

\begin{document}

\maketitle

\begin{abstract}
  We present results from a series of hydrodynamic simulations
  investigating ram pressure stripping of galactic haloes as the host
  galaxy falls radially into a cluster. We perform a parameter study
  comprising of variations in initial gas content, gas injection rate (via
  stellar mass loss processes), galaxy mass and amplitude of infall. From
  the simulation results we track variations in both physical quantities
  (e.g.\ gas mass) and directly observable quantities (e.g.\ X-ray
  luminosities). The luminosity of the galaxy's X-ray halo is found to
  compare favourably with the observationally determined correlation with
  optical blue band luminosity ($\rm{L_X:L_B}$) relation. Factors
  affecting the X-ray luminosity are explored and it is found that the gas
  injection rate is a dominant factor in determining the integrated
  luminosity. Observational properties of the material stripped from the
  galaxy, which forms an X-ray wake, are investigated and it is found that
  wakes are most visible around galaxies with a substantial initial gas
  content, during their first passage though the cluster. We define a
  statistical skewness measure which may be used to determine the direction
  of motion of a galaxy using X-ray observations. Structures formed in
  these simulations are similar to the cold fronts seen in observation of
  cluster mergers where a sharp increase in surface brightness is
  accompanied by a transition to a cooler region.

\end{abstract}


\section{Introduction}

Most galaxies are not isolated but reside in clusters or groups of galaxies
\citep{tully_1987} where a number of mechanisms may affect the nature and
evolution of the galaxy. These mechanisms include tidal interactions,
galaxy harassment, and ram pressure stripping. Ram pressure stripping,
which is the subject of this paper, has been identified in X-ray
observations of several nearby elliptical galaxies e.g.\ M86
(\citealt{rangarajan_1995}, \citealt{white_1991}, \citealt{forman_1979})
and NGC\,4472 \citep{irwin} in the Virgo Cluster and NGC\,1404 in the
Fornax cluster \citep{paolillo_2002}.  In these relatively nearby cases it
is possible to directly image the X-ray emission surrounding the galaxy,
which is distorted into an X-ray wake by the stripping process. Spectral
analysis can yield information on the temperature and metallicity structure
lending further weight to the hypothesis that the wake is the result of a
stripping process rather than an accretion process \citep{rangarajan_1995}.

To detect the effects of ram pressure stripping in more distant galaxies
requires a different approach. As wakes are often low surface brightness
features it is difficult to image wakes in distant systems, particularly
when viewed against a bright cluster background. If stripping affects
global properties, such as the total X-ray luminosity of the galaxy, this
could be detected, for example in an environmental dependence of the
$\rm{L_B}$ relation.  It is currently unclear how a galaxy's X-ray
luminosity $\rm{L_X}$ is affected by the cluster environment and, if so,
whether ram pressure is a significant or dominant mechanism. Different
studies of elliptical galaxies in clusters do not always draw the same
conclusions (e.g.\ \citealt{o'sullivan_2001,brown_2000}).  It is possible
that the nearby wake candidates are atypical and that ram pressure
stripping is not a common process. Alternatively, it may be the case that
$\rm{L_X}$ suppression is not a good indicator of the presence of stripping
or that other processes affect $\rm{L_X}$ more than ram pressure.

Numerical simulations of stripping processes provide valuable insights into
physical processes and their observable consequences which cannot be
obtained from analytical models alone. Previous work in this area includes
simulations of elliptical galaxies moving through a uniform cluster
environment at a constant velocity which is appropriate for galaxies on a
circular orbit within a cluster (e.g.\ \citealt{gaetz_1987,balsara_1994}).
Given the constant ram pressure in these cases a quasi-steady state is
reached and certain scaling relations (e.g.\ for retained gas mass fraction)
may be derived \citep{gaetz_1987}.  Although \cite{kent_1982} determine
that galaxy orbits in the Coma cluster cannot be primarily radial, the case
of a radial, or highly elliptical orbit, is still of considerable interest
in representing accretion of material from the field or a sub-cluster
merger.  NGC\,4839 is the dominant galaxy in a sub-cluster which is falling
into the Coma cluster and XMM-Newton observations (\citealt{neumann_2002},
\citealt{neumann_2000}) show an X-ray tail around NGC\,4839 which is
interpreted as a wake of stripped material.  As clusters grow via minor
mergers there may be many such cases of large elliptical galaxies passing
through clusters on radial paths.  \cite{takeda_1984} simulate radial
infall for a large ($10^{12} \rm{M}_\odot$) elliptical galaxy tracking the
galactic gas evolution for number of cluster crossings, finding a periodic
variation in gas content after the first cluster passage.  A detailed study
of elliptical galaxies on elliptical orbits has been carried out by
\cite{toniazzo_2001} using 3-dimensional simulations yielding detailed
information on the gas dynamic processes and variations in X-ray
luminosity.  A broad parameter study of stripping from dwarf galaxies has
been carried out by \cite{mori_2000} for galaxy masses in the range
$10^6$--$10^{10} \, \rm{M}_\odot$. Interactions between dwarf galaxies and
the cluster environment affect the chemical evolution of galaxies and
clusters; higher metallicity gas which is stripped from the galaxy will
enrich the lower metallicity medium of the cluster, and the depletion of
the gas content may prevent further star formation.  Ram pressure stripping
may also influence the morphology density relation \citep{dressler_1980}.

Our aim in the present study is to use 2-D hydrodynamic simulations of a
spherical galaxy falling radially into a cluster in order to determine the
effect of such an event on the galaxy and the impact on observable
properties such as the galaxy's X-ray luminosity.  A number of parameter
variations are investigated enabling determination of the sensitivity of
the results to the value of a given model parameter.  This parameter study
indicates not only how sensitive our results are to the choice of model,
but also what are the consequences of genuine variations in the physical
parameters of a galaxy population.  The aim of the parameter study
presented here is to focus on high mass galaxies where interactions with
the intra-cluster medium lead to more directly observable consequences than
in the lower mass cases considered by \cite{mori_2000}.  X-ray observations
of galaxy wakes caused by stripped material could be used to elucidate the
full 3-D nature of galaxy dynamics \citep{merrifield_1998} but a sample of
wakes observed in a given cluster will not be unbiased hence it is
important to understand the factors which influence which galaxies will
host observable wakes.

Section~\ref{section:model} presents the details of the galaxy model, the
model cluster into which the galaxy falls and the numerical scheme used.
Section~\ref{section:results} consists of a number of subsections which
present the results from the simulation parameter study.
Subsection~\ref{subsec:phases} gives an overview of the stripping phases
which occur during a cluster crossing, subsection~\ref{subsec:gas} examines
the variations of the galactic gas content, the observable consequences of
which are dealt with in subsection~\ref{subsec:halo}.
Subsection~\ref{subsec:wake} examines the properties of the wake of
stripped material and subsection~\ref{subsec:shock} explores the properties
of the bow shock.  The effect of placing the model galaxy in the harsher
stripping environment of a hotter cluster is examined in
section~\ref{sec:hotcl}.  Finally, section~\ref{section:conclusions}
presents a discussion of the results and conclusions which may be drawn.

\section{Overview of the model}
\label{section:model}

This section presents details of the canonical, or typical, simulation run.
The parameter study involves varying each parameter from its value in the
canonical model in order to determine the effect of altering a single model
component. For example, to investigate the effect of increasing the size of
the pre-existing gas halo, the initial extent of the halo was doubled
whilst all other parameters retained their standard values. When changing
the mass of the model galaxy, more than one simulation parameter was
altered. The mass of the dark matter halo was changed and other parameters
calculated so as to be consistent with the increased dark matter mass as
described below.

\subsection{The galaxy model}

The model galaxy consists of two distinct components; a stellar component
and a dark matter component. The dark matter component is considered to be
the more massive and hence dominates the gravitational influence of the
galaxy, however the stellar component is significant not only due to its
gravitational effects but also due to sources of gas injection which come
from stellar mass loss processes (e.g.\ type Ia supernovae, planetary
nebulae and stellar winds).  These injection processes place additional gas
onto the grid within the galaxy and can replenish a stripped gas halo.  A
model galaxy is constructed by specifying a mass for the dark matter halo
and then using a number of scaling relations to determine values for
subsequent parameters.

\subsubsection{Dark matter component}

Numerical simulations of structure formation by \cite{navarro_1997}
indicate that dark matter haloes have a universal cuspy density profile
where the dark matter density at a distance $r$ from the centre of the halo
is given by
\begin{equation}
  \label{eq:nfw_profile}
  \rho_{\rm{dm}}(r) = \frac{\rho_0}{cx \left( 1+cx \right)^2}
\end{equation}
where 
\begin{equation}
  \label{eq:x_dfn}
  x=\frac{r}{r_{200}},
\end{equation}
$c$ is the concentration parameter which governs the degree of central
concentration of the halo.  In hierarchical models of structure formation
smaller objects form earlier in the evolution of the universe and hence are
more centrally concentrated so the concentration parameter scales with halo
mass. The scaling relation used was that determined from the simulations of
\cite{bullock_2001},
\begin{equation}
  \label{eq:conc_para}
  c=10 \left( \frac{M_{\rm{halo}}}{2.1\times 10^{13} \, \rm{M}_\odot} 
\right)^{-0.14}.
\end{equation}
The value of $r_{200}$ depends on the redshift of formation of the object
and the matter density of the Universe at that redshift which in turn
depends on the assumed cosmology. Due to the difficulties in determining
these quantities a typical value of $r_{200}=50$ kpc was chosen for
$M_{\rm{halo}} = 2.0\times 10^{12} \, \rm{M}_\odot$ and it was assumed that
$r_{200} \propto M_{\rm{halo}}^{1/3}$.
The final constraint imposed on the dark matter halo
is that the dark matter distribution is truncated at $r_{200}$, so that
\begin{equation}
  \label{eq:dm_mass}
 M_{\rm{halo}} =
  \int^{r_{200}}_0 \rho_{\rm{dm}} \left( r \right) 4 \pi r^2 dr .
\end{equation}
\subsubsection{Stellar component}

Given the properties of the dark matter halo it is possible to define a
stellar population which is consistent with the observed properties of
elliptical galaxies. The total blue band luminosity of the stellar
population was determined assuming a total mass to light ratio of 50
$\rm{M}_\odot$/$\rm{L}_\odot$ based on Table~1 from
\cite{finoguenov_2000} (N.B. this is the mass to light ratio for the
whole galaxy, values measured using an aperture smaller than $r_{200}$ will
measure smaller mass to light ratios as dark matter dominates at large
radii). For the old stellar population of an elliptical galaxy the mass to
light ratio of the stars themselves will be greater than unity. The value
adopted was
\begin{equation}
  \label{eq:m/l_stars}
  \frac{M_{*}}{L_{*}} = 13 \frac{\rm{M}_\odot}{\rm{L}_\odot}
\end{equation}
based on an average value from the sample of \cite{finoguenov_2000} where
the original mass to light ratios were determined by \cite{lauer_1985}.
Given these assumed mass to light ratios the total stellar mass may then be
determined. Two mass to light ratios are required as the first dictates the
amount of dark matter relative to the galaxy's blue band luminosity and the
second describes the mass to light ratio of the stars only which depends on
the age of the stellar population.  The stars are assumed to be distributed
according to a King-type distribution as used by \cite{balsara_1994} i.e.
\begin{equation}
  \label{eq:rho_stars}
  \rho_* \left( r \right) = \frac{\rho_*
    \left(0\right)}{1+\left(r/R_{\rm{c}}\right)^2}
\end{equation}
where $R_{\rm{c}}$ is the core radius (2.5\,kpc) and the stellar
distribution is truncated at $R_{\rm{H}}=20 \, \rm{kpc}$. The central
stellar density $\rho_*\left(0\right)$ is determined using
\begin{equation}
  \label{eq:star_mass}
 M_* = \int^{R_{\rm{H}}}_0 \rho_* \left( r \right) 4 \pi r^2 dr .
\end{equation}
\subsubsection{Gas halo}

The pre-existing gas halo, which surrounds the galaxy at the start of the
simulation, is assumed to extend out to $r_{200}$ and be composed of
material of solar metallicity \citep{matsushita_2000}.  The halo is
initially isothermal at a temperature of 0.8 keV, the measured temperature
of the gas halo surrounding M86 \citep{matsushita_2000} which is of a
similar mass to our model galaxy. The effects of radiative cooling are
included which causes the dense central regions to cool appreciably from
the initial temperature.  The initial conditions are such that the halo is
in hydrostatic equilibrium with the galaxy potential and that there is
pressure continuity between the outer edge of the halo and the surrounding
intra-cluster medium (ICM).

\subsubsection{Mass replenishment and drop out}

Stellar mass loss processes (e.g\. type Ia supernova (SNIa), planetary
nebulae) will inject matter and energy into the galaxy which affects the
X-ray halo. The distribution of replenished material is assumed to follow
the stellar distribution so that the mass injection rate as a function of
radius is
\begin{equation}
  \label{eq:rhorep}
  \dot{\rho}_{\rm{rep}} = \alpha_* \frac{\rho_*\left(0\right)}
{1+\left(r/R_{\rm{c}}\right)^2}
\end{equation}
where $\alpha_*=5.4\times10^{-20}\, \rm{s}^{-1}$ is the specific mass loss
rate for an old stellar population as determined by \cite{brighenti_1998}.

The energy injection is dominated by SNIa and a component due to the
velocity dispersion of the stars in the galaxy potential. A supernova rate
typical of an elliptical galaxy is 0.1 SNU
(\citealt{finoguenov_2000,cappellaro_1993}) (1 SNU = 1 supernova per year
per $\rm{L_{B\odot}})$ and the typical energy injection from one SNIa is
$8\times 10^{50}$ ergs.

The velocity dispersion of the galaxy was estimated using the Faber-Jackson
relation
\begin{equation}
  \label{eq:faber-jackson}
  \frac{\rm{L_V}}{2\times10^{10} \, \rm{L}_\odot} = \left( \frac{\sigma}{200
\,\mbox{${\,\rm km~s}^{-1}$\,}
} \right)^4 
\end{equation}
\citep{faber_1976b} where $L_V$ was determined from the blue band
luminosity $\rm{L_B}$ assuming a $(B-V)$ colour of 0.95 \citep{bender_1993}.

In addition to processes which act as sources of gas there are other
processes which act as sinks to remove gas (e.g.\ star formation or
accretion onto a central compact object).  Radiative cooling is included,
using a Raymond-Smith thermal plasma model with solar metallicity, which
results in cool dense gas accumulating at the centre of the galaxy where it
is allowed to drop out of the flow and is removed over a given time-scale.
In a cell where the gas density is above the critical value of
$\rho_{\rm{do}}=2.4\times10^{-24} \, \rm{g} \, \rm{cm}^{-3}$ and the gas
temperature is below the critical value of $T_{\rm{do}}=2\times10^{4} \,
\rm{K}$ gas will be removed at a rate of
\begin{equation}
  \label{eq:dropout}
  \rho_{\rm{do}} = \frac{\rho}{\tau_{\rm{do}}}
\end{equation}
where $\tau_{\rm{do}}=3\times10^{7} \rm{yr}$.  The drop out prescription
used here is identical to that of \cite{stevens_1999}.

\subsection{Cluster model and galaxy dynamics}

\begin{figure*}
  \begin{center}
    \subfigure[Time variation of galaxy position relative to the cluster 
    centre. Positive values indicate that the galaxy is located to the
    right of the cluster centre and negative values indicate that the
    galaxy is located to the left.]{
    \includegraphics[scale=0.30,angle=-90]{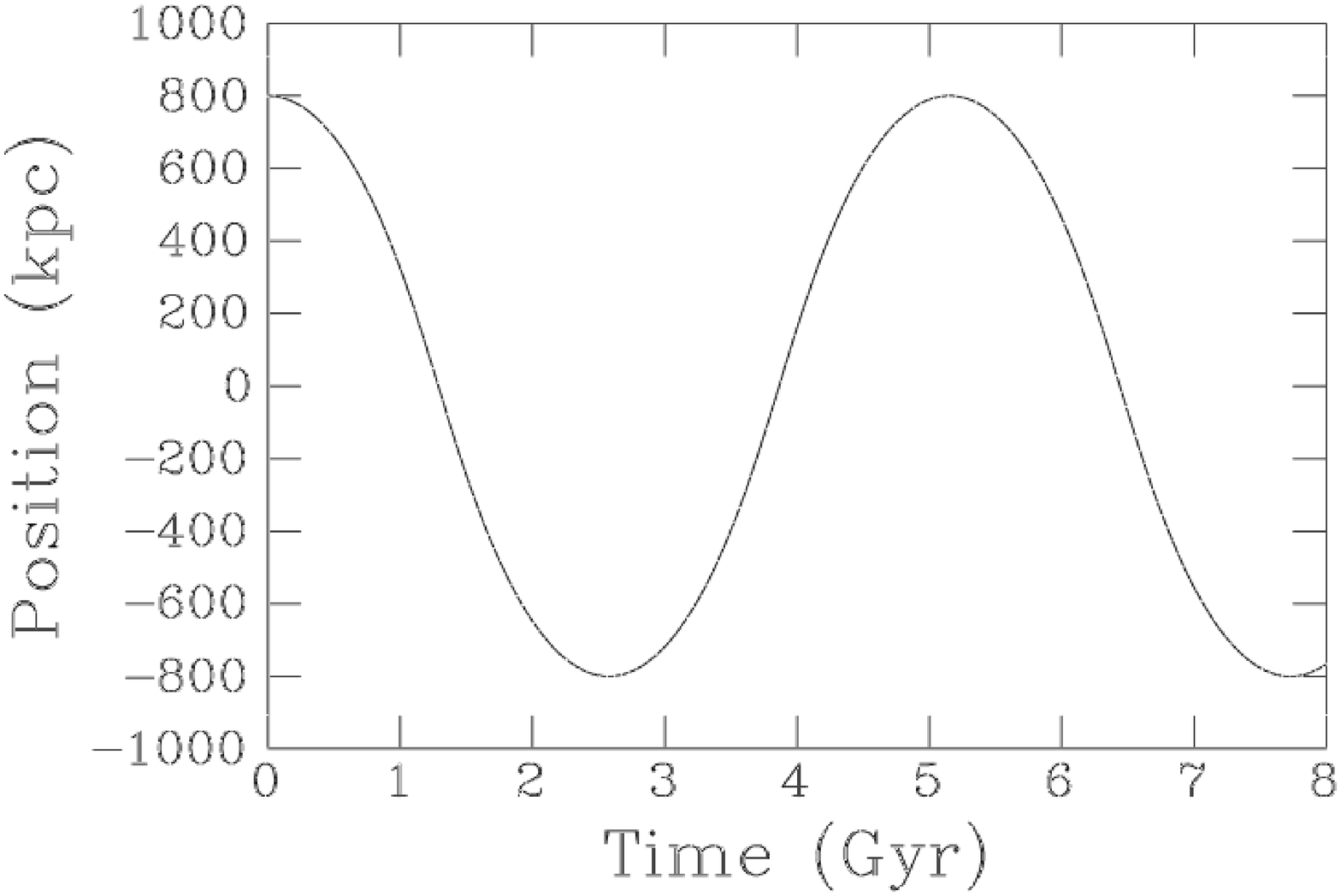}}
    \subfigure[Time variation of galaxy velocity. Horizontal lines mark the
    cluster sound speed of $847 \, \rm{km~s}^{-1}$]{
    \includegraphics[scale=0.30,angle=-90]{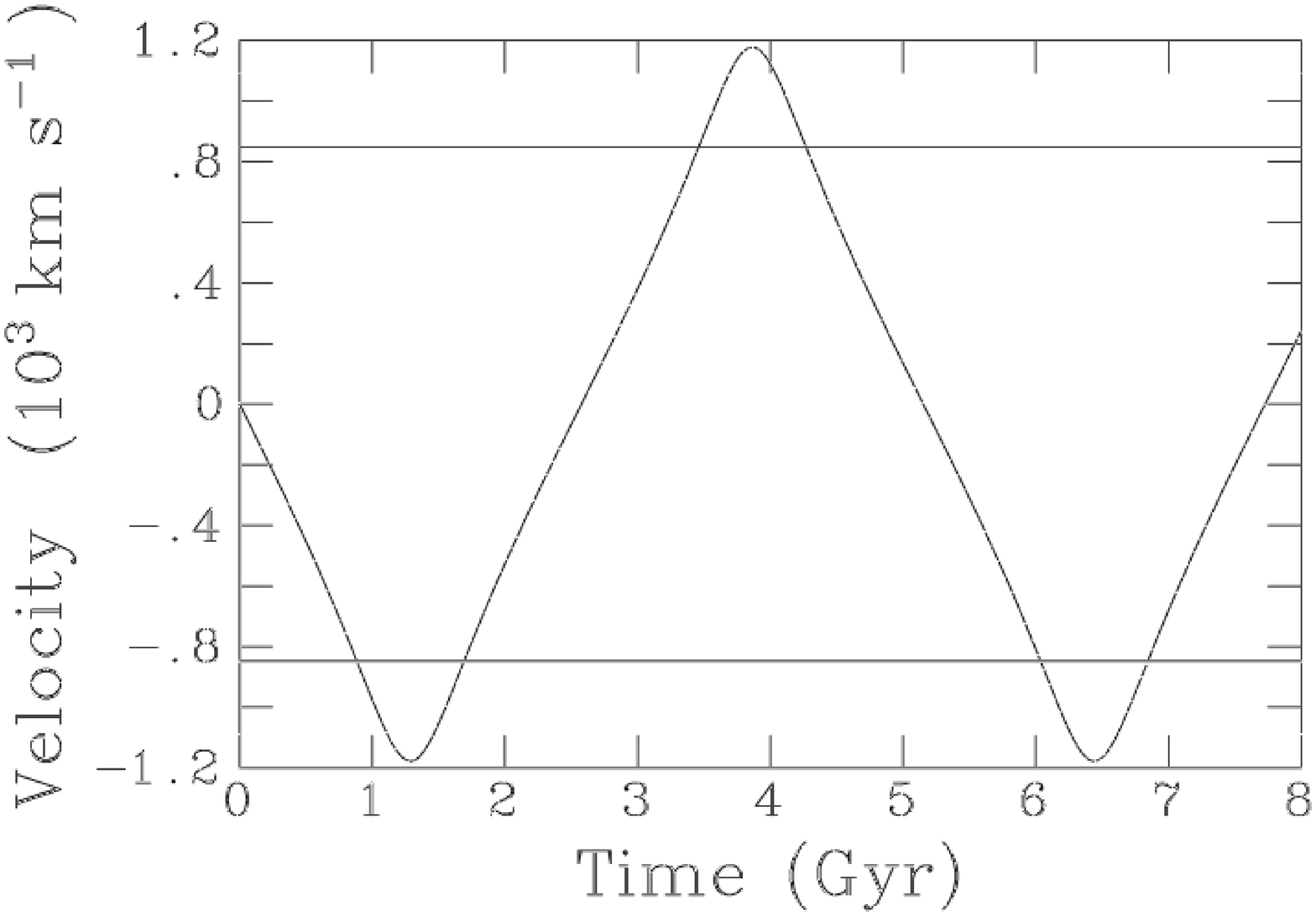}}
    \caption{Galaxy dynamics.}
    \label{fig:dyn}
  \end{center}
\end{figure*}
The model cluster was chosen to be a favourable environment for observing
wakes. In order for a significant number of wakes to be observed it is
required that the density of the ICM be high enough that ram pressure is
sufficient to distort the galaxy halo, but not so high that the background
emission from the cluster obliterates the signature of the wake. The
optimum range of cluster temperatures to fulfil these criteria is 2--3 keV
\citep{stevens_1999}.  The parameters of the cluster model were based on
observations of the cluster Abell\,160, a dynamically relaxed cluster at a
temperature of 2.7 keV and a distance of 270 Mpc, in which a statistical
detection of a number of wakes has been found \citep{drake_2000}.

The dynamics of the galaxy were modelled by free-fall in the cluster
potential, neglecting dissipation processes such as dynamical friction but
including the effect of tidal forces.  The form of the cluster potential
was chosen to be the same as that used by \cite{toniazzo_2001}
\begin{equation}
  \label{eq:clust_pot}
  \phi \left(r\right) = \sigma_{\rm{c}}^2 \ln \left[ 1 + 
  \left(\frac{r}{r_{\rm{c}}} \right)^2 \right]
\end{equation}
where $\sigma_{\rm{c}}$ is the cluster velocity dispersion and $r_{\rm{c}}$
is the cluster core radius.  The free fall frame of reference in which the
simulation is conducted transforms away first order gravitational effects
but not tidal effects.  For a potential of this form, tidal forces are
greatest at the cluster core radius.  A core radius of 400 kpc was adopted
to match the value of \cite{white_1997} for Abell 160. The galaxy was
allowed to fall into the cluster, on a radial trajectory, from an initial
position of 2 cluster core radii from the centre and an initial state of
rest, resulting in the dynamics shown in Fig.~\ref{fig:dyn}.  The shape of
the cluster potential is fixed by eqn.~\ref{eq:clust_pot} but the depth of
the potential, parameterized by the cluster velocity dispersion, is not
fixed. The X-ray surface brightness $S(w)$ at a projected distance $w$ from
the centre of a relaxed cluster is found to be well fitted by a beta model
with $\beta=2/3$
\begin{equation}
  \label{eq:beta-model}
  S(w)=\frac{S_0}{\left[ 1+ \left( w/r_c \right)^2 \right] ^{3/2}}
\end{equation}
which results from a gas density profile of the form
\begin{equation}
  \label{eq:gas_density_profile}
  \rho = \frac{\rho_0}{1+ \left(r/r_{\rm{c}} \right)^2}
\end{equation}
Gas in hydrostatic equilibrium in the potential given in
eqn.~\ref{eq:clust_pot} will have the density distribution given in
eqn.~\ref{eq:gas_density_profile} if
\begin{equation}
  \label{eq:sigma}
  \sigma_{\rm{c}}^2 = \frac{kT}{\mu m_{\rm{H}}}
\end{equation}
where $\mu m_{\rm{H}}$ is the mean mass per ICM particle.  A value of
$\sigma_{\rm{c}} = 656 \, \rm{km~s}^{-1}$ was used, derived from
eqn.~\ref{eq:sigma} with a temperature of 2.7 keV.  The central ICM number
density was determined by relating the total X-ray luminosity $\rm{L_X}$
and the central surface brightness $S_0$ for the surface brightness
profile given in eqn.~\ref{eq:beta-model}
\begin{equation}
  \label{eq:lx_sb}
   L_X = 8 \pi^2 S_0 r^{2}_{\rm{c}}.
\end{equation}
The central surface brightness may be related to the central emissivity and
hence to central number density. Assuming a plasma composed of hydrogen and
helium in the ratio 10:1 the total central number density was estimated to
be $6.17 \times 10^{-4} \,\rm{cm}^{-3}$ for $\rm{L_X} = 2.7 \times 10^{43}
\, \rm{erg~s}^{-1}$ \citep{white_1997}.

\subsection{Numerical scheme}
\label{num_method}

The numerical scheme chosen for this application was the VH-1
implementation of the Piecewise-Parabolic Method \citep{colella_1984} which
offers good resolution of shocks in complex flows and has been extensively
used in a variety of astrophysical contexts (\citealt{blondin_1990},
\citealt{richards_1998} and \citealt{strickland_2000}). The numerical grid
was a uniform Cartesian grid of $800\times400$ cells with a size of
$x=1.6\times 10^{24}\;\rm{cm}$, $y=8.0\times 10^{24}\;\rm{cm}$ where the
centre of the galaxy was located at $x=8.0\times 10^{24}$, $y=0$. In an
axi-symmetric case, such as this, it is only necessary to simulate the top
half of the galaxy and the symmetry of the situation can be used to reflect
the flow field in the $x$-axis to obtain the full flow field which results in
a significant saving in computation time. The size of the grid as a whole
was chosen such that features of interest, such as tails of stripped
material and shocks, should not propagate off the grid. As the galaxy moves
both from right to left and from left to right during the course of the
simulation it is necessary to centre the galaxy with respect to the $x$-axis.
The cells subtend an angle of $\sim0.5\,\rm{arcsec}$ at a distance of
$270\,\rm{Mpc}$ (the distance of Abell 160) which matches the spatial
resolution of the Chandra X-ray observatory, currently the best available
X-ray spatial resolution.

Boundary conditions at the left and right of the grid were fixed inflow or
fixed outflow conditions to account for the movement of the simulation
frame in the gravitational potential of the model cluster (e.g.\ fixed
inflow at the left boundary and fixed outflow at the right boundary for a
galaxy moving from right to left).  The magnitude of the inflow velocity at
one boundary was equal to the magnitude of the outflow velocity at the
other boundary so that the direction of motion could be reversed in a
consistent fashion.  The fixed inflow/outflow densities were different due
to density gradients in the ICM. The size of the grid is not negligible
compared to the size of the cluster so the ICM density is different at the
left and right boundaries. To ensure a physically consistent model for the
cluster, it is necessary to ensure the ICM contains a density gradient
consistent with hydrostatic equlibrium in cluster potential given by
eqn.~\ref{eq:clust_pot}. If eqn.~\ref{eq:sigma} is obeyed then the
resulting X-ray surface brightness distribution of the ICM will be given by
a beta model with $\beta=2/3$ as in eqn.~\ref{eq:beta-model}. This density
gradient must be present in the initial ICM distribution and must also
be accounted for when calculating inflow and outflow densities at the left
and right boundaries.

A reflective boundary condition must be
applied to the $y=0$ boundary in order for the symmetry of the problem to
be correctly applied and an inflow/outflow condition was applied to the top
of the grid to allow material to flow freely off the grid should it reach
the upper boundary. Time steps were determined by application of the
Courant-Friedrichs-Lewy (CFL) condition with a CFL number of 0.8.

The use of a 2D numerical grid results in a significantly shorter
computation time than for a 3D grid, for comparable resolution, which is an
important issue when performing the many runs required to cover a wide range
of parameter space. The rate of stripping may depend on the formation of
Kelvin-Helmholtz shear instabilities which can behave differently in 2D and
3D simulations. \cite{bruggen_2001a} compare the mixing rate due to such
instabilites in 2D and 3D simulations and find that the steady state mixing
rate is very similar, supporting the validity of the 2D treatment used
here. It should be noted, however, that if magnetic field effects are
included there is significant disagreement between 2D and 3D simulations
\citep{bruggen_2001b}. As no magnetic fields are explicitly included in
these simulations we proceed with a 2D model.

\section{Results}
\label{section:results}

In this section results are presented from the parameter study examining
both observational quantities, such as X-ray luminosity and surface
brightness profiles, and other quantities (e.g.\ gas mass) which,
whilst not directly observable, play an important role in determining the
nature of the galaxy.  The parameters involved in the study were: galaxy
mass, mass injection rate, initial gas halo extent and the starting
position of the galaxy (hence the maximum velocity reached) as shown in
Table~\ref{tab:par_study}.
\begin{table*}
  \begin{center}
    \caption{Details of runs forming the parameter study}
    \begin{tabular}{lccc}
\hline
Parameter         & Low Value & Standard Value & High value\\
\hline
Halo extent             & 25 kpc & 50kpc & 100 kpc              \\
Mass replenishment rate & $0.12 \, \rm{M}_\odot~\rm{yr}^{-1}$& $1.2  \, 
\rm{M}_\odot~\rm{yr}^{-1}$& $3.6 \, \rm{M}_\odot~\rm{yr}^{-1}$ \\
Dark matter halo  mass  & $1.0 \times 10^{12}\,\rm{M_\odot}$ &
$2.0\times10^{12}\,\rm{M_\odot}$ & $  4.0 \times 10^{12}\,\rm{M_\odot}$\\
Infall radius             &400kpc  & 800 kpc & 1600  kpc            \\
\hline 
    \end{tabular}
    \label{tab:par_study}
  \end{center}
\end{table*}

\subsection{Initial cluster crossing}
\label{subsec:phases}
\begin{figure*}
  \begin{center}
    \subfigure[$t=0\,\rm{Gyr}$, $d=800\,\rm{kpc}$]{
    \label{fig:dens_0}
    \includegraphics[scale=0.3,angle=-90]{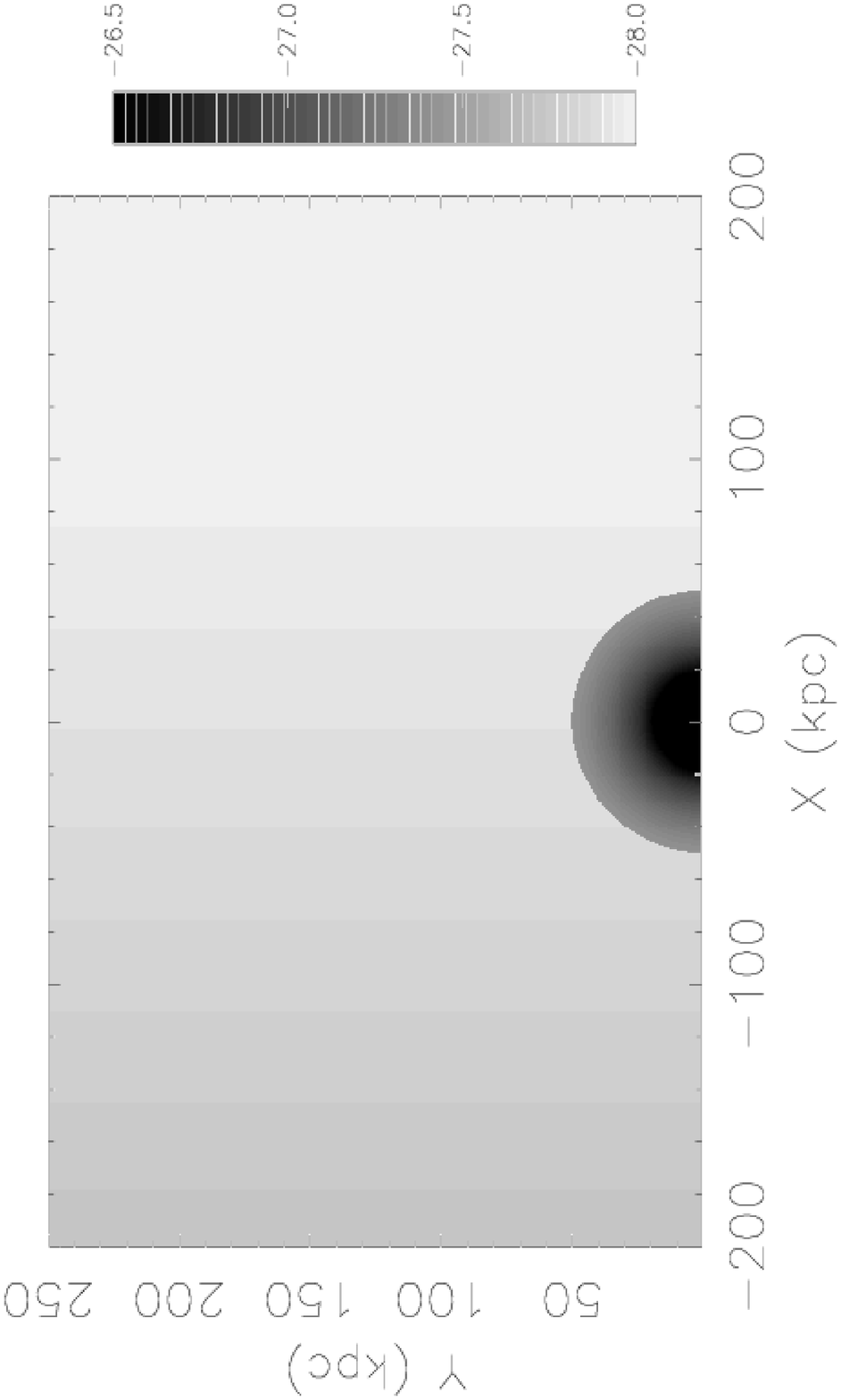}}
    \subfigure[$t=0.8\,\rm{Gyr}$, $d=500\,\rm{kpc}$]{
    \label{fig:dens_0.8}
    \includegraphics[scale=0.3,angle=-90]{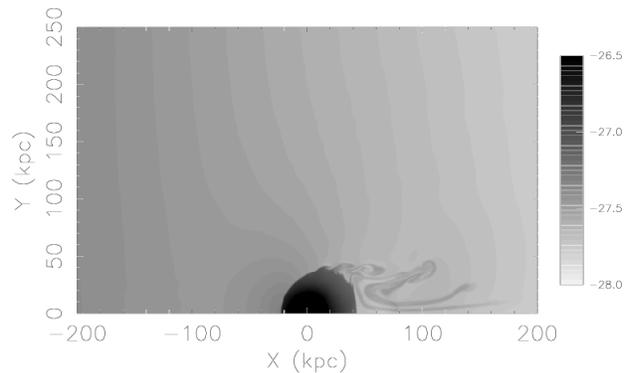}}
    \subfigure[$t=1.6\,\rm{Gyr}$, $d=-350\,\rm{kpc}$]{
    \label{fig:dens_1.6}
    \includegraphics[scale=0.3,angle=-90]{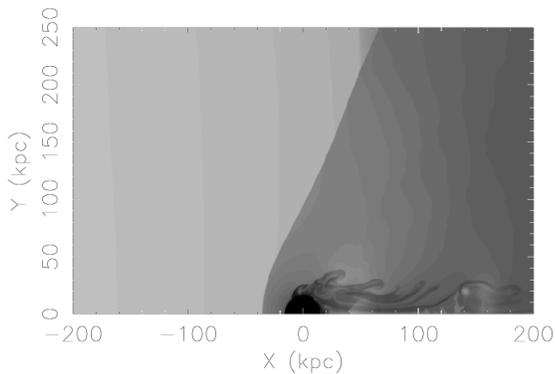}}
    \subfigure[$t=2.4\,\rm{Gyr}$, $d=-790\,\rm{kpc}$]{
    \label{fig:dens_2.4}
    \includegraphics[scale=0.3,angle=-90]{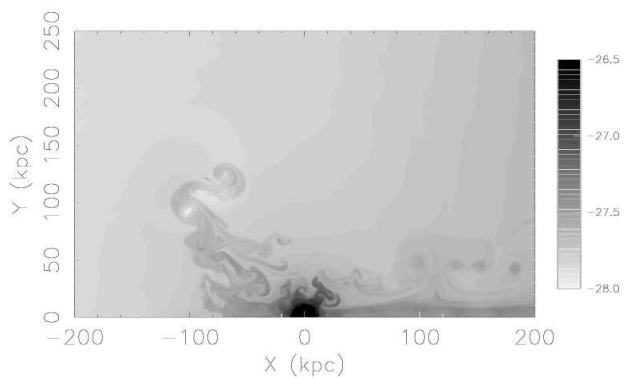}}
    \caption{Gas density distribution from the initial infall of the
      canonical run. The time ($t$) quoted is that since the start of the
      simulation and $d$ is the distance from the cluster centre which is
      positive when the galaxy is to the right of the cluster centre and
      negative when to the left. The logarithmic greyscale shows the gas
      density in units of $\rm{g\,cm}^{-3}$.}
    \label{fig:dens}
  \end{center}
\end{figure*}

Density distributions from the canonical run are shown in
Fig.~\ref{fig:dens} to illustrate various phases of the stripping process.
Fig.~\ref{fig:dens_0} is the initial gas distribution (i.e.\ a
spherically symmetric halo in hydrostatic support) which is beginning to be
stripped in Fig.~\ref{fig:dens_0.8} where the motion is subsonic. The halo
presents a sharp edge in the upstream direction but stripping, via
Kelvin-Helmholtz instabilities, prevents a sharp delineation in other
regions.  Fig.~\ref{fig:dens_1.6} is taken from a supersonic regime and a
bow shock, and associated density enhancement, can be seen. The final frame
(Fig.~\ref{fig:dens_2.4}) is taken at the end of the first cluster crossing
and density enhancements are seen in front of the galaxy. As the galaxy
approaches the edge of the cluster, and the ram pressure decreases,
material which was previously confined behind the galaxy moves forward
causing the enhancements seen in Fig.~\ref{fig:dens_2.4}.

\subsection{Gas content}
\label{subsec:gas}

\begin{figure*}
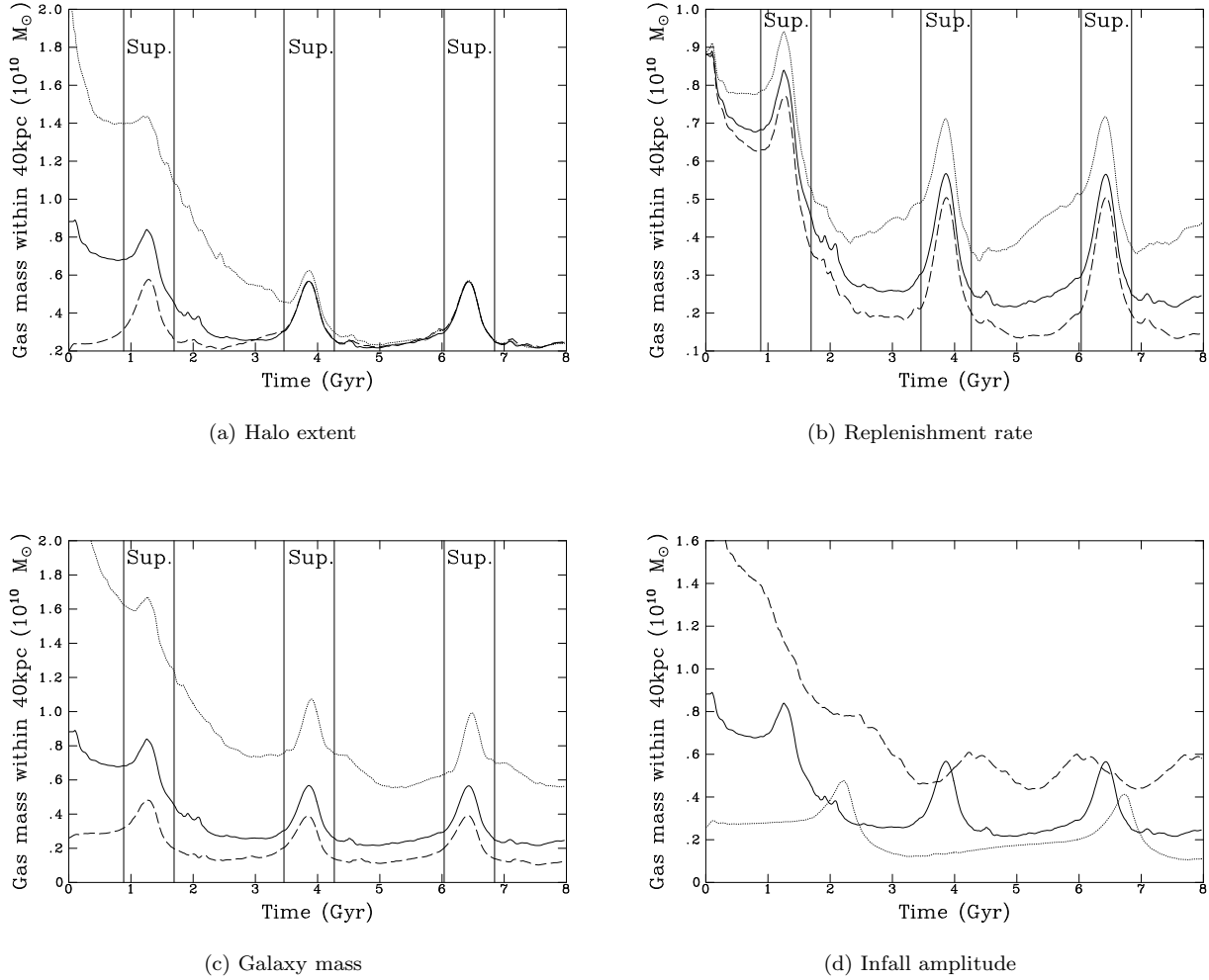

  \begin{center}
    \subfigure[Halo extent]{
      \label{fig:mass_halo}
      \includegraphics[scale=0.30,angle=-90]{fig3a.ps}}
    \subfigure[Replenishment rate]{
      \label{fig:mass_rep}
      \includegraphics[scale=0.30,angle=-90]{fig3b.ps}}
    \subfigure[Galaxy mass]{
      \label{fig:mass_mass}
      \includegraphics[scale=0.30,angle=-90]{fig3c.ps}}
    \subfigure[Infall amplitude]{
      \label{fig:mass_infall}
      \includegraphics[scale=0.30,angle=-90]{fig3d.ps}} 
    \caption{Time variation of the gas mass within 40 kpc of the galactic
      centre. In each case the canonical run is plotted as a solid line,
      the dashed line represents a reduction in the parameter and the
      dotted line an increase in the value of the parameter. Time periods
      where the galaxy velocity is supersonic are marked ``Sup.''. Periods
      of supersonic velocity are not shown in Fig.~\ref{fig:mass_infall} as
      they are different for each case.}
    \label{fig:mass}
  \end{center}
\end{figure*}
\begin{figure*}
  \centering
  \subfigure[Halo extent]{
  \label{fig:boundgas_halo}
  \includegraphics[scale=0.3, angle=-90]{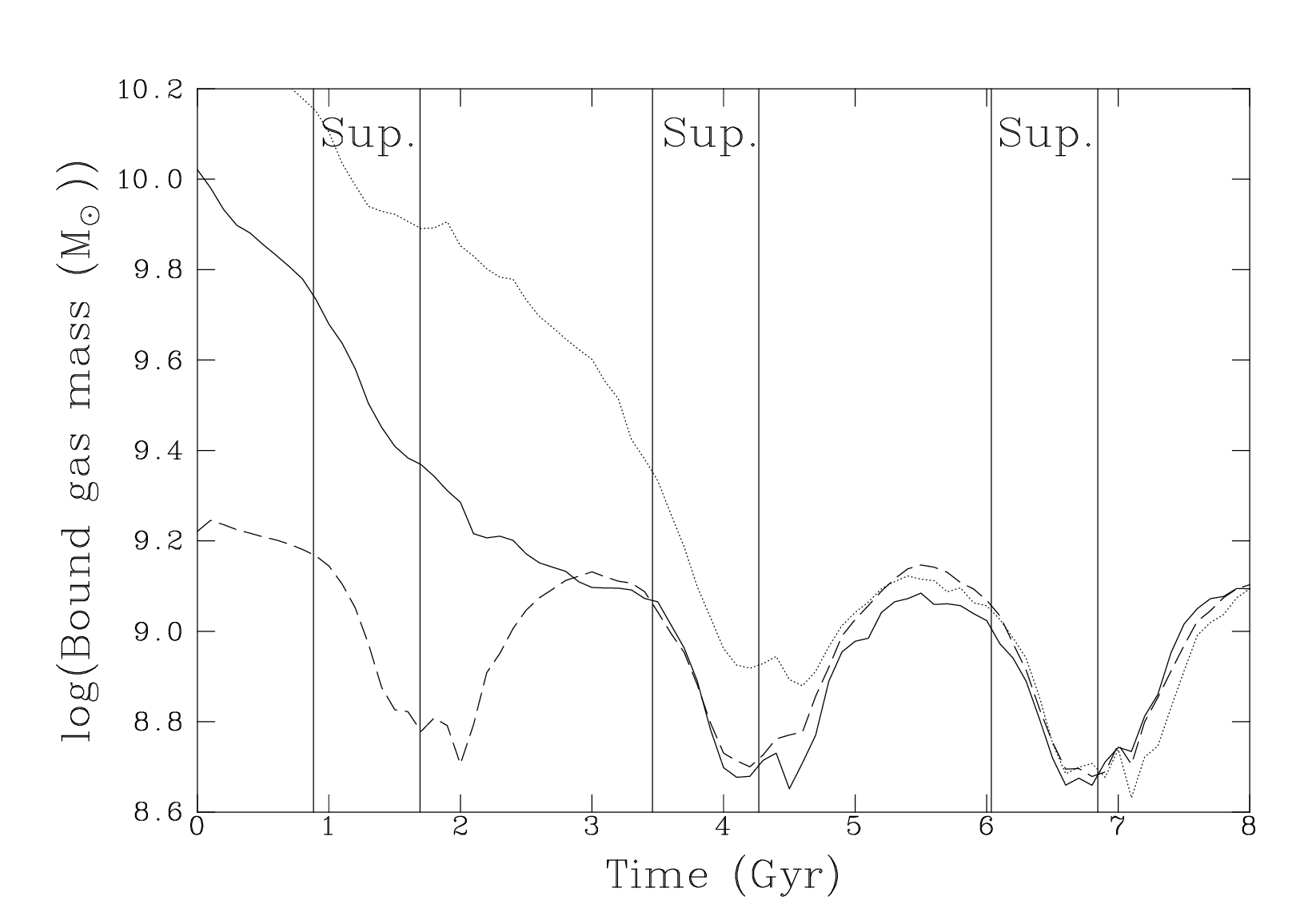}}
  \subfigure[Replenishment rate]{
  \label{fig:boundgas_mrep}
  \includegraphics[scale=0.3, angle=-90]{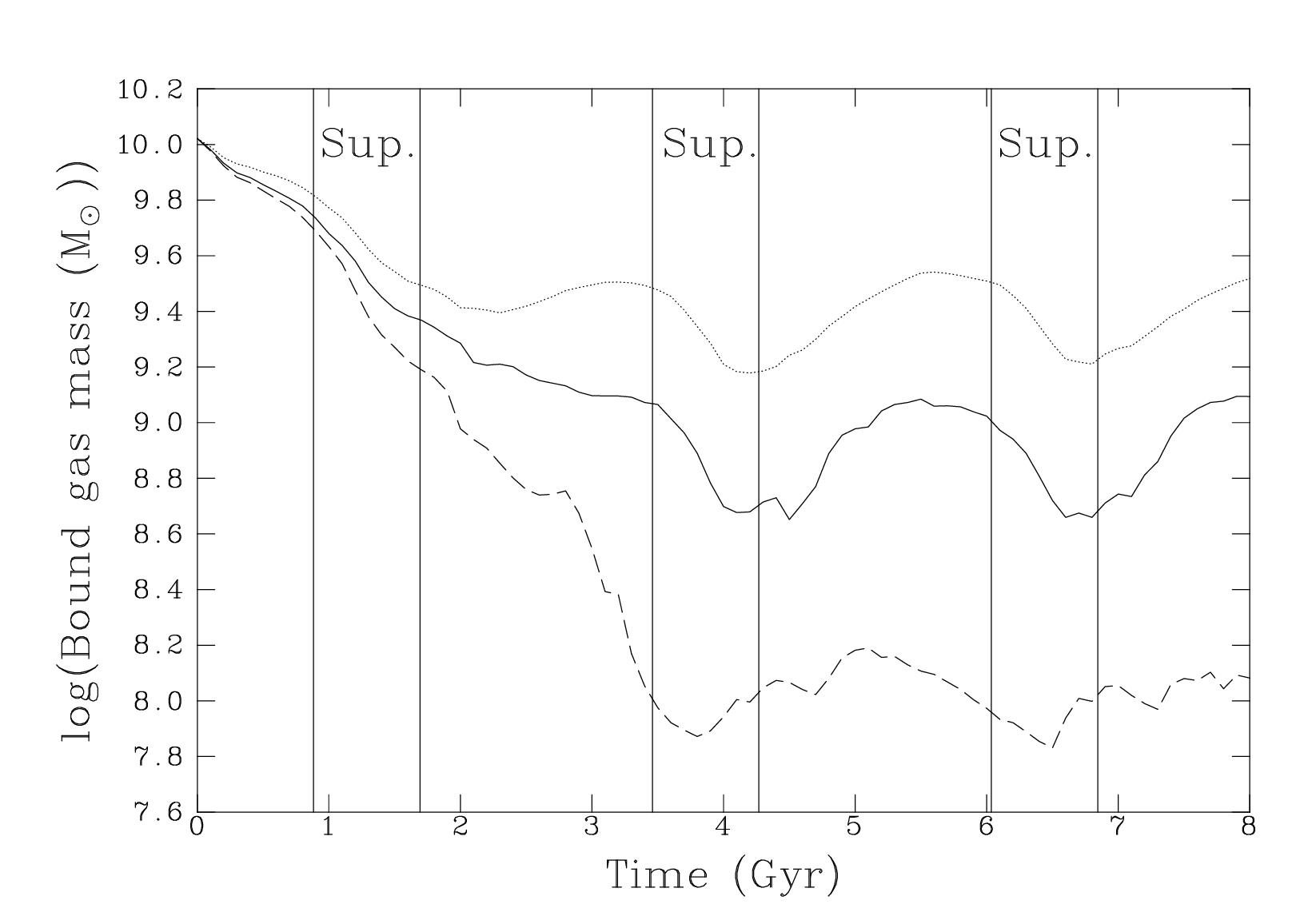}}
  \subfigure[Galaxy Mass]{
  \label{fig:boundgas_mass}
  \includegraphics[scale=0.3, angle=-90]{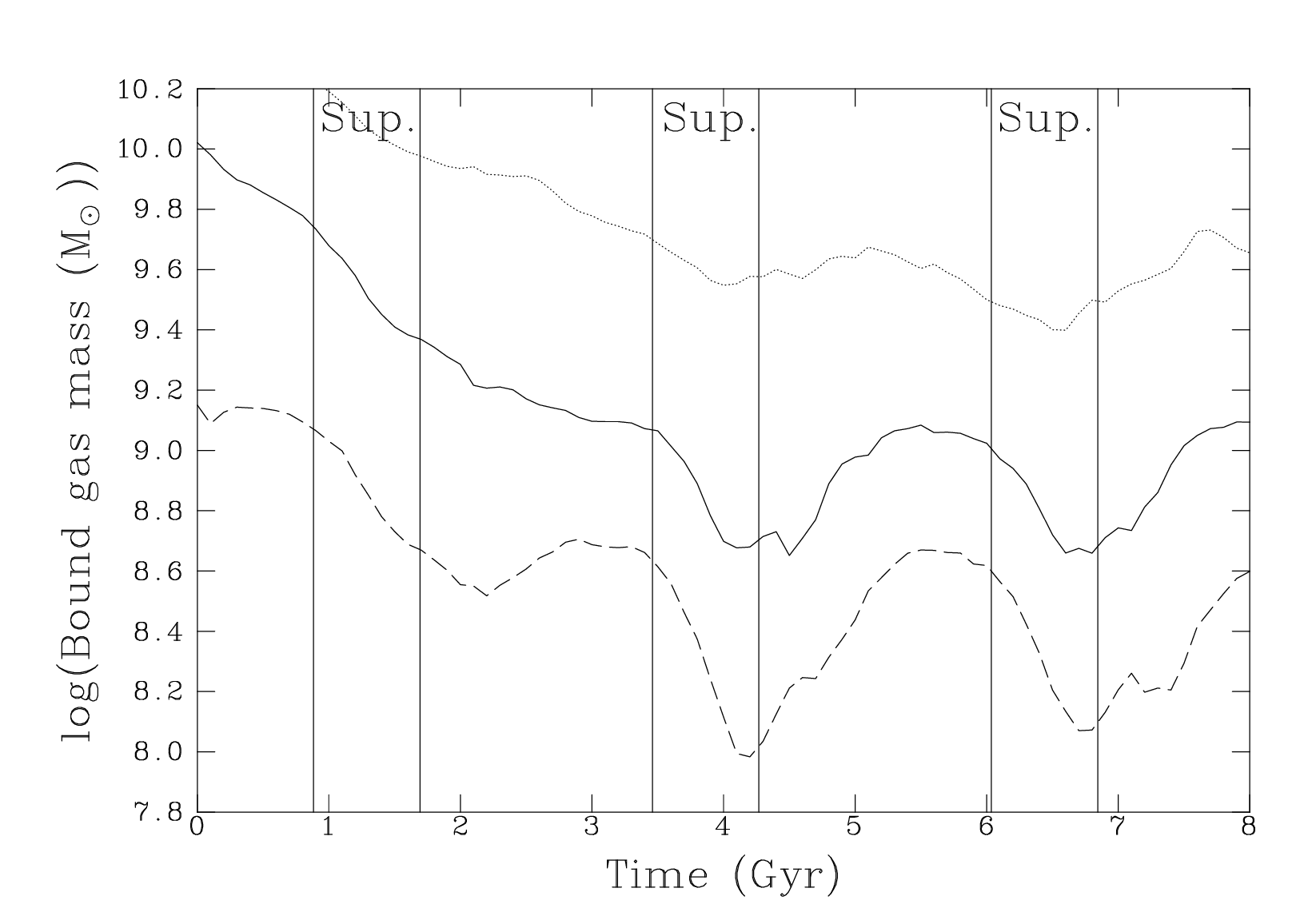}}
  \subfigure[Infall amplitude]{
  \label{fig:boundgas_infall}
  \includegraphics[scale=0.3, angle=-90]{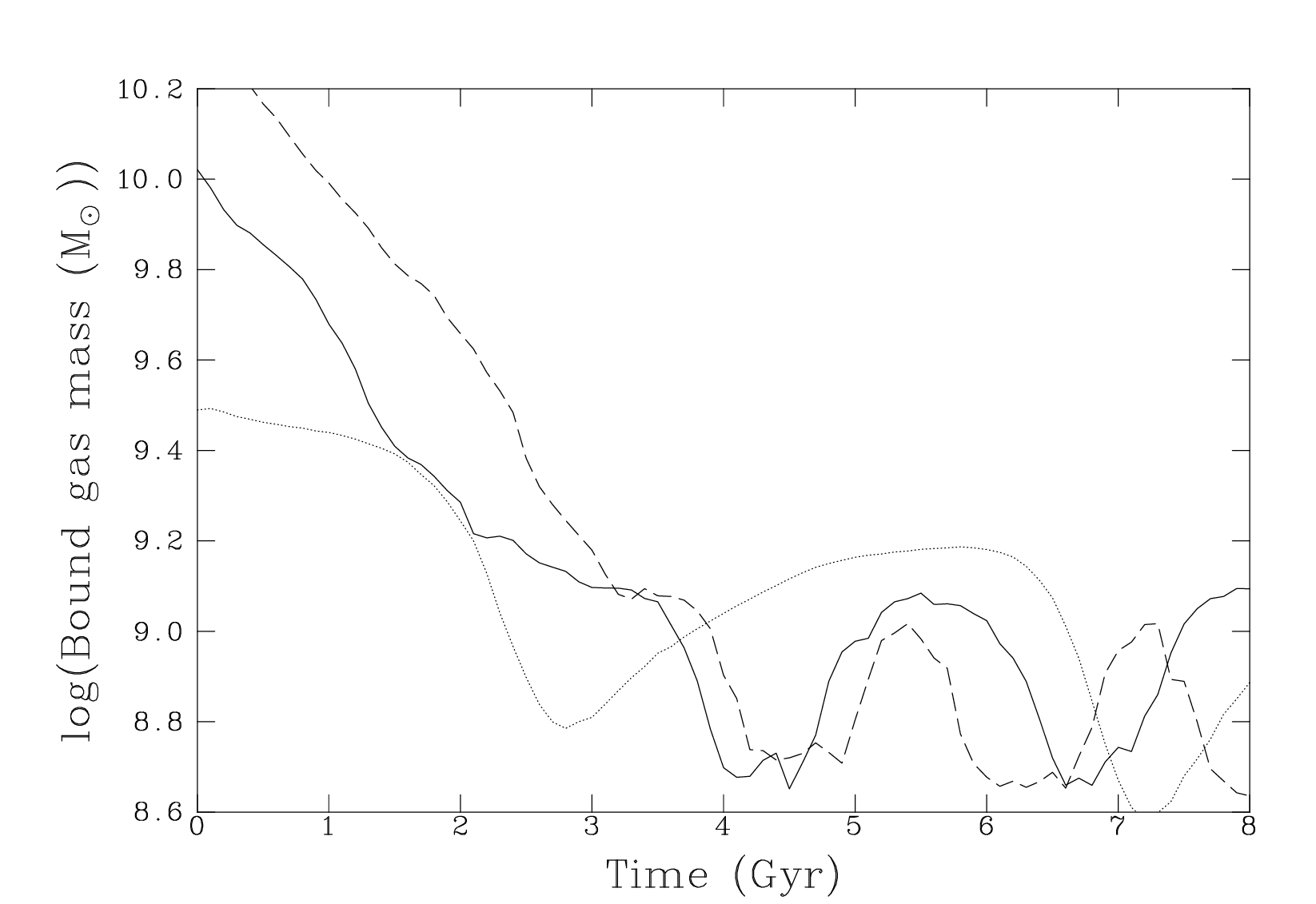}}
  \caption{Total gas mass which is gravitationally bound to the galaxy 
    (c.f.\ Fig.~\ref{fig:mass} which plots all gas within a fixed 40 kpc
    radius).  In each case the canonical run is plotted as a solid line,
    the dashed line represents a reduction in the parameter and the dotted
    line an increase in the value of the parameter. }
  \label{fig:boundgas}
\end{figure*}
\begin{figure}
  \centering
  \includegraphics[scale=0.4,angle=0]{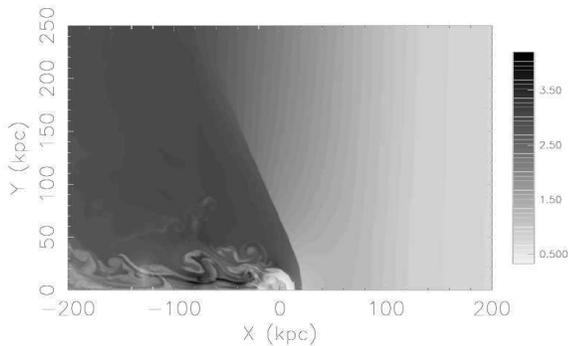}
  \caption{Gas density field after 3.9 Gyr divided by gas density field
    after 3.6 Gyr showing the location of the density enhancement around the
    time of core passage. The galaxy is moving from left to right at both
    times.}
  \label{fig:den_ratio}
\end{figure}

The variation in gas mass within the central 40 kpc, for galaxies with
different initial halo extents, is plotted in Fig.~\ref{fig:mass_halo}.
The vertical lines mark regions where the galaxy velocity is supersonic
(indicated by Sup.) close to core passage.  A more extensive gas halo is
more massive not only due to its larger volume but also due to the higher
central density which is necessary to satisfy the conditions of hydrostatic
support and pressure continuity at the halo truncation radius.  Both the
canonical and extended haloes experience an initial period during which the
pre-existing halo is stripped before moving into a cyclic phase where the
enclosed gas mass is at a maximum at core passage. The duration of the
initial stripping phase depends on the halo mass at the start of the
simulation ($\sim 2.5$ Gyr for the canonical halo and $\sim 4.0$ Gyr for
the larger halo) but after this phase all runs behave in a very similar
fashion indicating that after $\sim 4$ Gyr (time of second core passage)
the signatures of the initial gas content are no longer present. The large
enclosed mass close to core passage is somewhat surprising, but one should
bear in mind that this gas is not necessarily gravitationally bound to the
galaxy. Fig.~\ref{fig:boundgas_halo} shows the mass of gas which is
gravitationally bound to the galaxy for the three cases considered above.
The gas in a cell where the mass density is $\rho$, the particle number
density is $n$, the fluid velocity $u$ and the temperature $T$ is
considered to be bound if
\begin{equation}
  \label{eq:bound}
  \frac{1}{2}\rho u^2 + \frac{3}{2} nkT + \rho \psi_G < 0 
\end{equation}
where $\psi_G$ is the gravitational potential of the galaxy evaluated at
the cell location. 
In contrast to Fig.~\ref{fig:mass_halo}, which plots all gas within 40 kpc
irrespective of whether it is bound, Fig.~\ref{fig:boundgas_halo} indicates
that the supersonic phase results in a substantial amount of the bound gas
being stripped. This indicates that around the time of core passage there
is a concentration of gas within 40 kpc which is not gravitationally bound
to the galaxy and the question of where this gas is located is raised.
Fig.~\ref{fig:den_ratio} plots the ratio of gas density after 3.9 and 3.6
Gyr, i.e.\   the gas density at one of the peaks in Fig.~\ref{fig:mass_halo}
divided by the gas density just after the peak starts to form. This
indicates that the rise in unbound gas mass within 40 kpc is associated
with an enhancement of material behind the shock front.
When the halo extent is reduced by half, the initial period of stripping
prior to the cyclic behaviour is not seen as the initial gas content does
not exceed that present during the latter phases as in the other cases.

The gas mass for runs with different mass replenishment rates is shown in
Fig.~\ref{fig:mass_rep}.  In the previous series of runs the initial gas
content was different but the injection rate was constant resulting in the
evolution of the gas content in the model galaxies converging. In this
second case the initial gas content is constant but the injection rate
varies, resulting in a gas content which displays an initial period of
divergence. At later times (after $\sim 3$ Gyr) the gas content again
behaves in a cyclic manner but with a base level which depends on
replenishment rate.  The rate at which the mass increases prior to the
supersonic phase is a function of the mass replenishment rate with higher
injection rates resulting in more rapid rises in gas content. The higher
injection rate run shows how in subsonic phases there is an accumulation of
gas, during the supersonic phase the gas content is high due to the
accumulation of unbound gas, and after the supersonic phase much of the gas
associated with the galaxy has been depleted and is replenished during the
next subsonic phase. The bound gas content (Fig.~\ref{fig:boundgas_mrep}) 
better shows the effect of alternate cycles of net gas accumulation during
subsonic phases and net stripping during supersonic and some subsonic
regimes. When the mass injection rate is at its lowest it is seen that the
net stripping phase starts sooner than in other runs. During subsonic
motion, stripping does occur from the outer edge of the halo but the net
gas budget is a combination of the stripping and drop out loss mechanisms, 
and the mass injection replenishment mechanism.  Whether there is net loss
or gain of gas during a subsonic phase will depend largely on the
effectiveness of subsonic stripping compared to the mass injection rate.

\begin{figure}
  \begin{center}
    \includegraphics[scale=0.30,angle=-90]{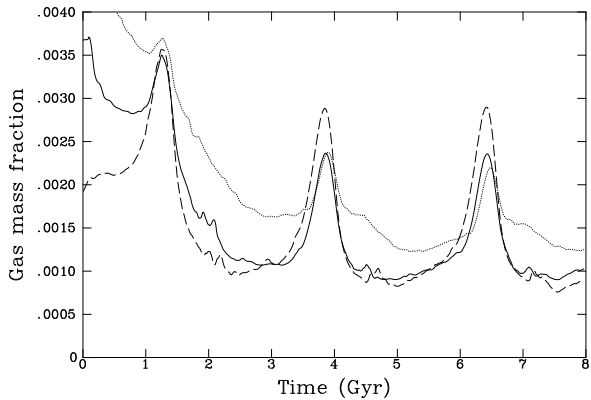}
    \caption{Gas mass fraction within 40 kpc of the galactic
centre for galaxies with differing masses (solid: canonical, dashed: less
massive, dotted: more massive). This is essentially Fig.~\ref{fig:mass}
with the $y$-axis divided by the total mass of the galaxy. }
    \label{fig:gasfraction}
  \end{center}
\end{figure}

The next series of runs involves variation of the galaxy mass which is
achieved by changing the dark matter halo mass and assuming length scales
vary as $r\propto M^{1/3}$.  By varying the total galaxy mass, not only is
the depth of the potential well increased but also the mass replenishment
rate (which is proportional to the stellar mass), the mass replenishment
temperature (which depends on the stellar velocity dispersion) and the
extent of the initial halo (which is subject to the same linear scaling
factor as the other galaxy length scales). The resulting interdependency of
parameters means that more care is required in interpreting the results of
these particular simulation runs but gives a self consistent galaxy model.
The gas mass fraction, within a 40 kpc radius, for each galaxy is plotted
in Fig.~\ref{fig:gasfraction} where the gas mass fraction has been
calculated by dividing the gas mass within 40 kpc by the total galaxy mass
(dark matter and stellar components).  The gas mass fraction indicates that
the more massive galaxies have a higher gas content in the subsonic regime.
The enhancement due to the bow shock, during supersonic phases, is greater
for less massive galaxies.  As the concentration parameter scales inversely
with galaxy mass, the less massive galaxies will be more centrally
concentrated and this may result in a more prominent shock.
The bound gas content (Fig.~\ref{fig:boundgas_mass}) indicates that more
massive galaxies not only have more gravitationally bound gas but also 
the variation in this gas content is less. The deeper gravitational
potential wells of the more massive systems would be expected to make
depletion of gas by ram pressure stripping less effective.

\begin{figure}
  \begin{center}
    \includegraphics[scale=0.30,angle=-90]{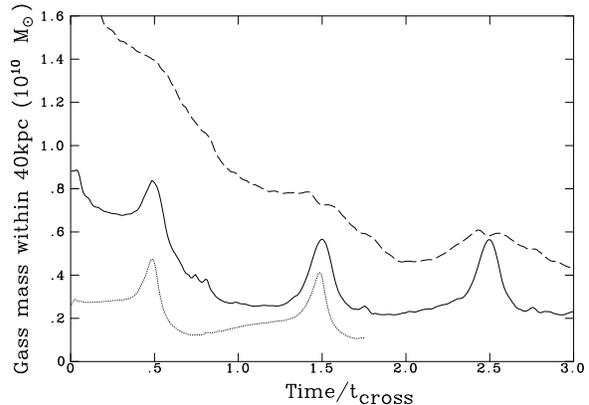}
    \caption{Time variation of the gas mass within 40 kpc of the galactic
centre for galaxies falling into the cluster from different starting
positions with the time axis scaled to the cluster crossing time 
(solid: canonical, dashed: infall from closer to the cluster
centre, dotted: infall from further out from cluster centre).}
    \label{fig:mass_infall_tcross}
  \end{center}
\end{figure}

The final parameter to be varied is the initial starting position from
which the galaxy falls into the cluster. This affects the galaxy dynamics
(i.e.\ the maximum speed reached by the galaxy and also the time taken to
cross the cluster, hence the periodicity in mass variation).  When the
infall radius is reduced to 400~kpc the maximum velocity reached is
770~$\rm{km\,s^{-1}}$ and the crossing time is $\sim1.7~\rm{Gyr}$; in this
case the galaxy velocity never becomes supersonic. An increased infall
radius of 1600~kpc raises the maximum velocity to 1560~$\rm{km\,s^{-1}}$
with a crossing time of $\sim4.5~\rm{Gyr}$.  The initial state of the gas
halo also varies in these runs. When the starting position of the galaxy is
closer to the cluster core, where the ICM density is higher, the halo
density must also be higher to maintain pressure continuity at the halo
boundary and hydrostatic support. The time variation of the gas mass for
these runs, shown in Fig.~\ref{fig:mass_infall}, clearly indicates the
different dynamical time-scales: with a larger infall the peaks at core
passage are $\sim4.5$ Gyr apart, with a smaller infall the peaks in mass
are separated by $\sim1.7$ Gyr. With a larger infall there is a marked
difference in gas mass before and after the peak at core passage suggesting
that substantial stripping occurs during the supersonic phase but that this
is masked by the density enhancement from the unbound gas. The gas mass
variation, with the time axis is scaled to the cluster crossing time
(Fig.~\ref{fig:mass_infall_tcross}), shows that the variations in gas mass
occur at similar points during the dynamic cycle.  The bound gas variations
for the above series of runs are plotted in Fig.~\ref{fig:boundgas_infall}.
The canonical and smaller infall cases have a similar minimum gas content
but with a smaller infall amplitude the excursions in gas content are
smaller.

Analytic fits to simulation results have been derived by \cite{gaetz_1987}
to determine the fraction of replenished gas which is retained in a
quasi-steady state, but it should be noted that the stripping occurring in
these simulations operates in a different regime. The results of
\cite{gaetz_1987} are applicable during supersonic regimes only (Mach.
numbers greater than unity), which applies only briefly in these
simulations, and in cases where the escape velocity from the galaxy
(measured at the half mass replenishment radius) exceeds the velocity of
the galaxy through the ICM. In the cases considered here, for the canonical
dynamics, the escape velocity is comparable with the maximum galaxy
velocity hence the fitting formulae of \cite{gaetz_1987} do not, and would
not be expected to, give a good fit to our results.

\subsection{Observable halo properties}
\label{subsec:halo}

\begin{figure*}
    \subfigure[Halo extent]{
    \label{fig:lx_vs_t_halo}
    \includegraphics[scale=0.3, angle=-90]{fig8a.ps}}
    \subfigure[Replenishment rate]{
    \label{fig:lx_vs_t_rep}
    \includegraphics[scale=0.3, angle=-90]{fig8b.ps}}
    \subfigure[Galaxy mass]{
    \label{fig:lx_vs_t_mass}
    \includegraphics[scale=0.3, angle=-90]{fig8c.ps}}
    \subfigure[Infall amplitude]{
    \label{fig:lx_vs_t_infall}
    \includegraphics[scale=0.3, angle=-90]{fig8d.ps}}
    \label{fig:lx_vs_t}
    \caption{Time evolution of X-ray luminosity ($\rm{L_X}$) in the 
      0.3--8.0 keV band for each parameter variation. In each case the
      canonical run is plotted as a solid line, the dashed line represents
      a reduction in the parameter and the dotted line an increase in the
      value of the parameter. }
\end{figure*}
\begin{figure}
  \begin{center}
    \includegraphics[scale=0.3, angle =-90]{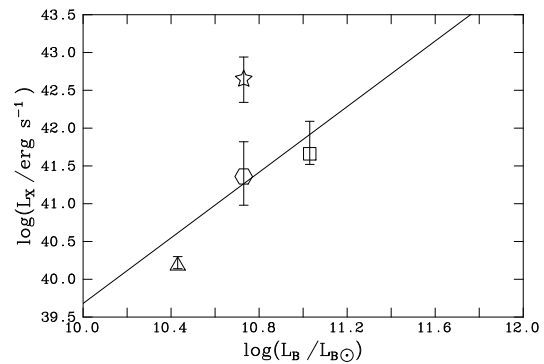}
    \caption{$\rm{L_X}$:$\rm{L_B}$ relation for four of the simulated
      galaxies (less massive (triangle), more massive (square), canonical
      mass with lower mass replenishment rate (hexagon) and canonical mass
      with higher mass replenishment rate (star)). $\rm{L_X}$ is the X-ray
      luminosity in the 0.5--2.0\,keV band and $\rm{L_B}$ is the blue band
      optical luminosity in solar units.  The solid line is the
      observationally determined relation of O'Sullivan et al. (2001). Bars
    in the $\rm{L_X}$ direction show the range in $\rm{L_X}$ during the
    simulation (see text for details).}
    \label{fig:lxlb}
  \end{center}
\end{figure}
The time variation of X-ray luminosity within a radius of 40 kpc, in the
0.3--8.0 keV band (i.e.\ a broad band Chandra or XMM luminosity), for runs
with varying initial halo extents is shown in Fig.~\ref{fig:lx_vs_t_halo}.
These luminosities only include the contribution from the gas halo and no
point source contribution.  The total X-ray luminosity is seen to broadly
follow the variations in gas mass within this fixed aperture, i.e.\ peaks at
core passage and converges to a common pattern of fluctuations when the
halo has been stripped.

The time variation of the galaxy X-ray luminosity for different values of
the mass replenishment rate is shown in Fig.~\ref{fig:lx_vs_t_rep}. 
The mass replenishment rate has a large impact on the X-ray
luminosity; the canonical and higher mass replenishment rate runs 
differ only by a factor of 3 in mass replenishment rate but have very
different X-ray luminosities.

The time variation of the galaxy X-ray luminosity, for different galaxy
masses, is shown in Fig. \ref{fig:lx_vs_t_mass}.  The least massive galaxy
has a significantly lower X-ray luminosity than the canonical case whereas,
in contrast, the most massive galaxy has a similar luminosity to the
canonical galaxy. The X-ray luminosity is largely governed by the central
regions of the gas halo so differences in gas mass may be present which do
not manifest themselves as significantly different X-ray luminosities.

The time variation of galaxy X-ray luminosity, for different infall radii,
is shown in Fig.~\ref{fig:lx_vs_t_infall}. The different cluster crossing
times result in different periodicities in the $\rm{L_X}$ variations but it
is not possible to distinguish between the different dynamical states based
on the value of $\rm{L_X}$ alone.  An $\rm{L_X}$:$\rm{L_B}$ relation for
four of the simulated galaxies (lower galaxy mass, higher galaxy mass,
canonical mass with lower mass replenishment rate, canonical mass with
higher mass replenishment rate) is presented in Fig.~\ref{fig:lxlb}.  The
upper and lower bounds indicate the maximum and minimum $\rm{L_X}$ reached
during simulation times between 2 and 8 Gyr (to avoid any initial transient
effects) and the symbols indicate the mean $\rm{L_X}$ measured during the
same period.  The lower mass replenishment rate run is responsible for the
lowest $\rm{L_X}$ for a galaxy of canonical mass, and the higher mass
replenishment rate run is responsible for the highest $\rm{L_X}$ hence the
inclusion of these two points indicates the range of $\rm{L_X}$ which is
induced by parameter variations.  The X-ray luminosities between 2 and 8
Gyr have been converted from a broad band (0.3--8.0 keV) to a soft band
(0.5--2.0 keV) to enable a direct comparison between X-ray luminosities
determined from ROSAT and the theoretical values derived here.  A point
source contribution, which scales with $\rm{L_B}$, has been added assuming
\begin{equation}
  \label{eq:lxlb}
  \log\left(\frac{L_{\rm{ps}}}{L_{\rm{B}}}\right) = 29.45
\end{equation}
where $\rm{L_{ps}}$ is the contribution from point sources only
(\citealt{ciotti_1991,o'sullivan_2001}).  The X-ray luminosities derived
for these simulated systems are in agreement with the observational
relation of \cite{o'sullivan_2001} which is plotted as a solid line in
Fig.~\ref{fig:lxlb}.  Variations in $\rm{L_X}$ due to observing a given
model galaxy at varying points in the dynamical cycle are smaller than
those variations induced by differing model parameters. The largest
variation for an individual model is for the model with a larger initial
halo extent which experiences $\rm{L_X}$ variations of $\sim1.4$ dex,
however the effect of parameter variations (particularly mass replenishment
rate) induces a scatter of $\sim2$ dex indicating that variations in $\rm{L_X}$
are not solely due to observing galaxies at different points in the
dynamical cycle but rather that there are more fundamental differences
causing $\rm{L_X}$ variations.

\begin{figure*}
  \begin{center}
    \subfigure[Surface brightness profiles during the initial infall, after
    0.8 (solid), 1.6 (dashed) , 2.4 (dotted) Gyr. Surface brightness is
    measured in units of
    $\rm{erg}\,\rm{s}^{-1}\,\rm{cm}^{-2}\,\rm{arcmin}^{-2}.$]
    {\includegraphics[scale=0.30,angle=-90]{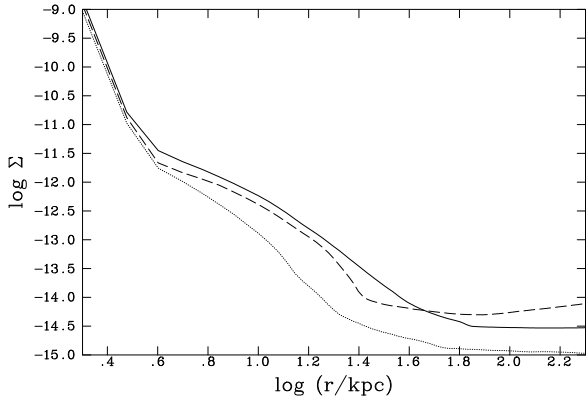}}
    \subfigure[Surface brightness profiles during the second cluster
    crossing, after 3.4 (solid), 4.2 (dashed) , 5.0 (dotted) Gyr.  Surface
    brightness is measured in units of
    $\rm{erg}\,\rm{s}^{-1}\,\rm{cm}^{-2}\,\rm{arcmin}^{-2}.$]
    {\includegraphics[scale=0.30,angle=-90]{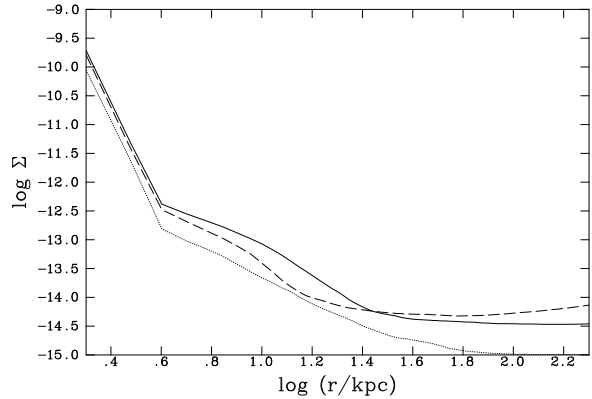}}
    \subfigure[Temperature profiles during the initial infall, after 0.8
    Gyr (solid), 1.6 Gyr (dashed), 2.4 (dotted) Gyr.]
    {\label{fig:temp_prof_init}
      \includegraphics[scale=0.30,angle=-90]{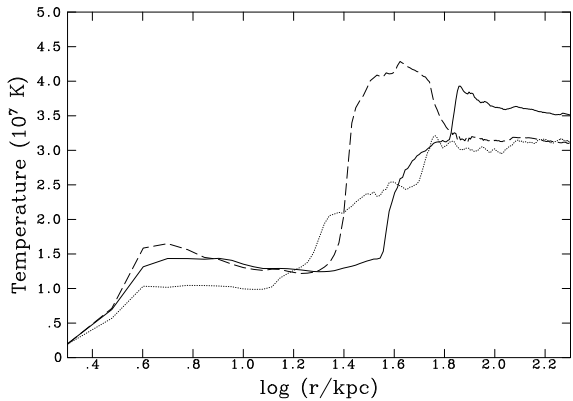}}
    \subfigure[Temperature profiles during second crossing, after 3.4 Gyr
    (solid), 4.2 Gyr (dashed), 5.0 (dotted) Gyr.]
    {\label{temp_prof_second}
      \includegraphics[scale=0.30,angle=-90]{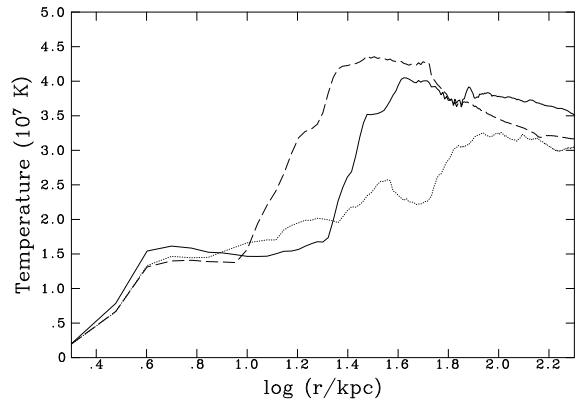}}
    \caption{Azimuthally averaged radial surface brightness profiles and 
      temperature profiles from the canonical run. The centre of the galaxy
      is located at $r=0$. Surface brightness $\Sigma$ is in units of 
      $\rm{erg}\,\rm{s}^{-1}\,\rm{cm}^{-2}\,\rm{arcmin}^{-2}.$}
    \label{fig:surface_brightness_profiles}
  \end{center}
\end{figure*} 
\begin{figure*}
  \centering
  \subfigure[Surface brightness after 0.8 Gyr in units of 
  $\rm{erg}\,\rm{s}^{-1}\,\rm{cm}^{-2}\,\rm{arcmin}^{-2}.$ The galaxy
  velocity is subsonic at this time.]
  {\label{fig:sb_0.8}
  \includegraphics[scale=0.30, angle=-90]{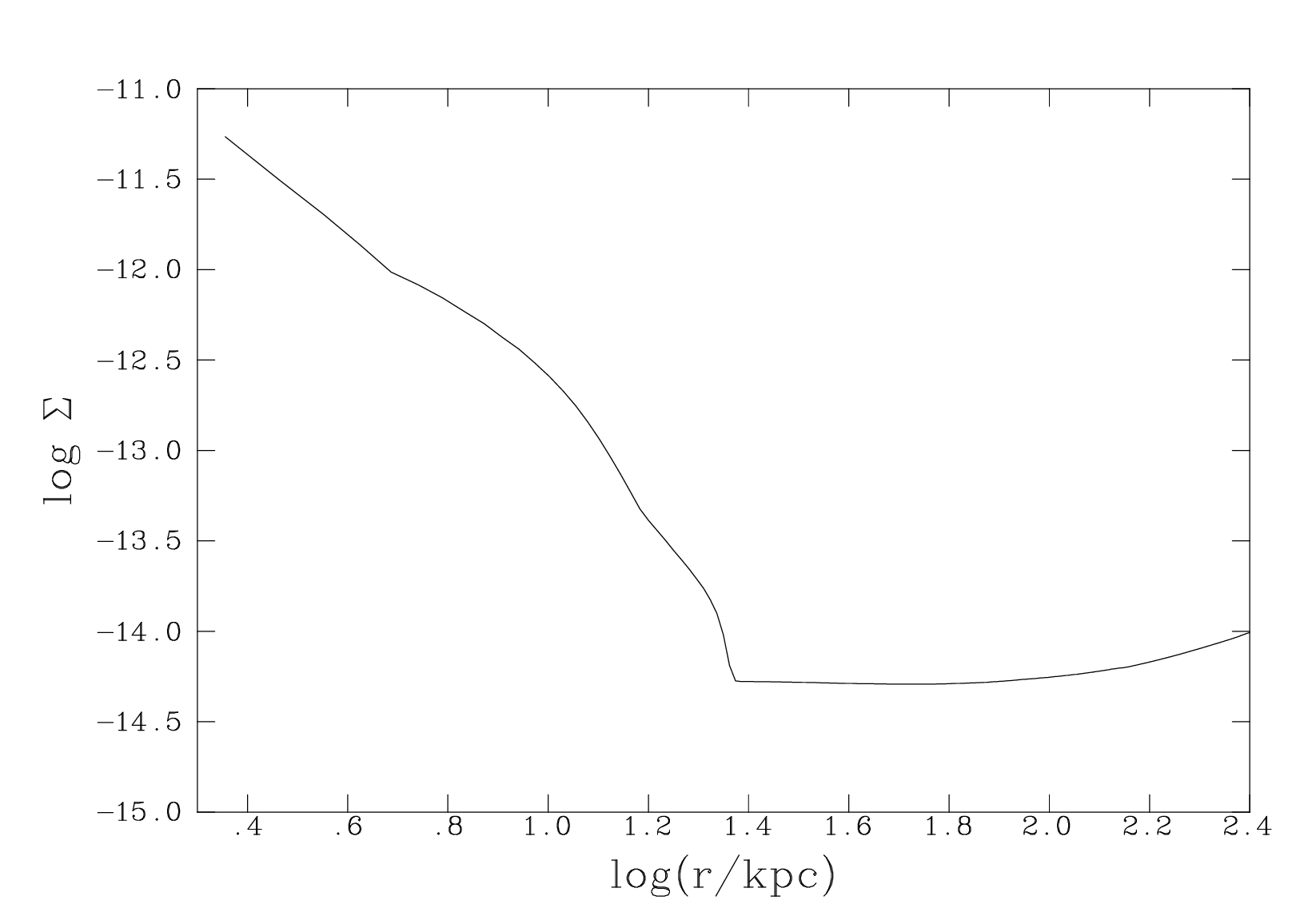}}
  \subfigure[Surface brightness after 1.6 Gyr in units of 
  $\rm{erg}\,\rm{s}^{-1}\,\rm{cm}^{-2}\,\rm{arcmin}^{-2}.$ The galaxy
  velocity is supersonic at this time.]
  {\label{fig:sb_1.6}
  \includegraphics[scale=0.30, angle=-90]{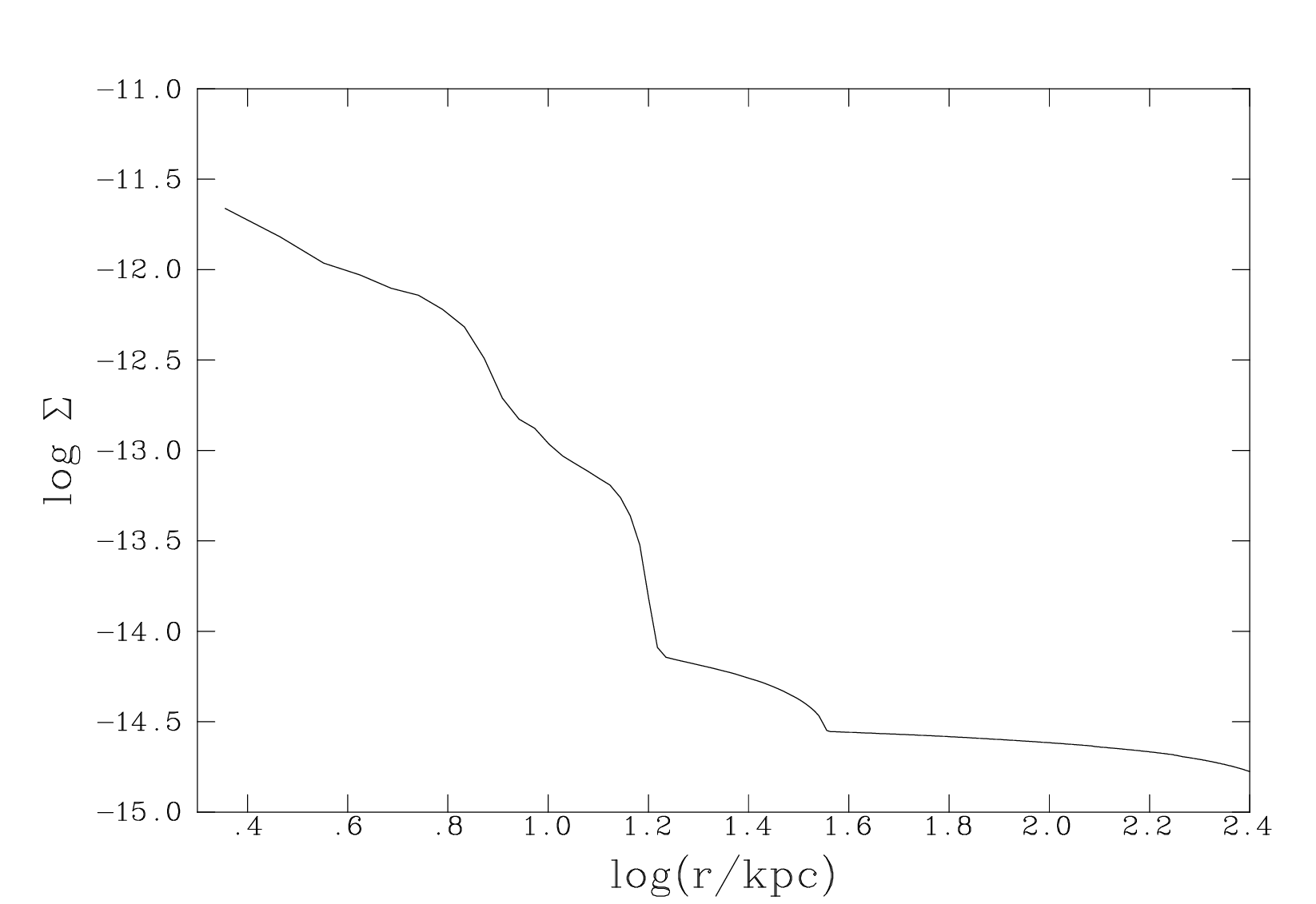}}
  \subfigure[Temperature profile after 0.8 Gyr. The galaxy
  velocity is subsonic at this time.]
  {\label{temp_0.8}
  \includegraphics[scale=0.30, angle=-90]{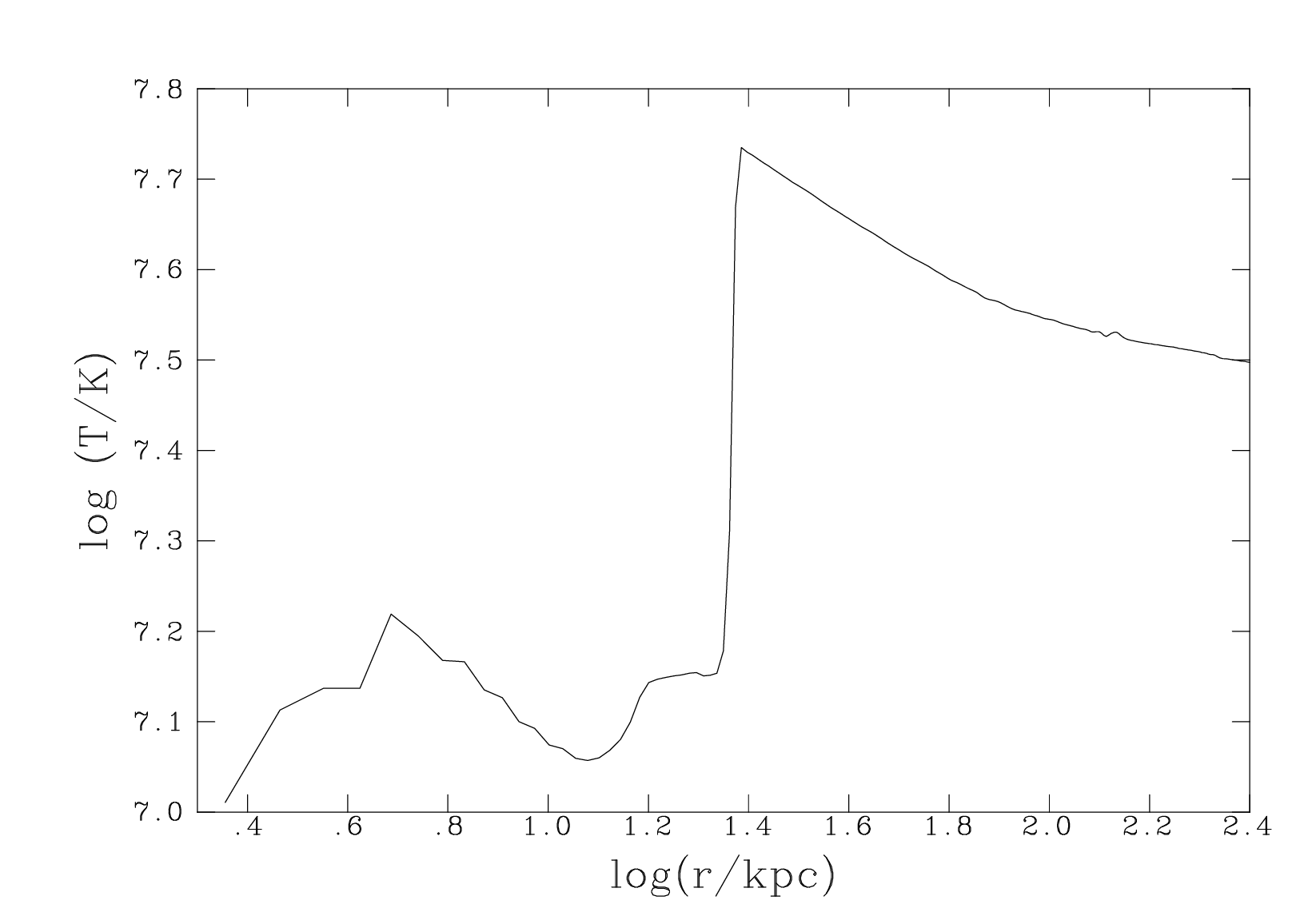}}
  \subfigure[Temperature profile after 1.6 Gyr. The galaxy
  velocity is supersonic at this time.]
  {\label{temp_1.6}
  \includegraphics[scale=0.30, angle=-90]{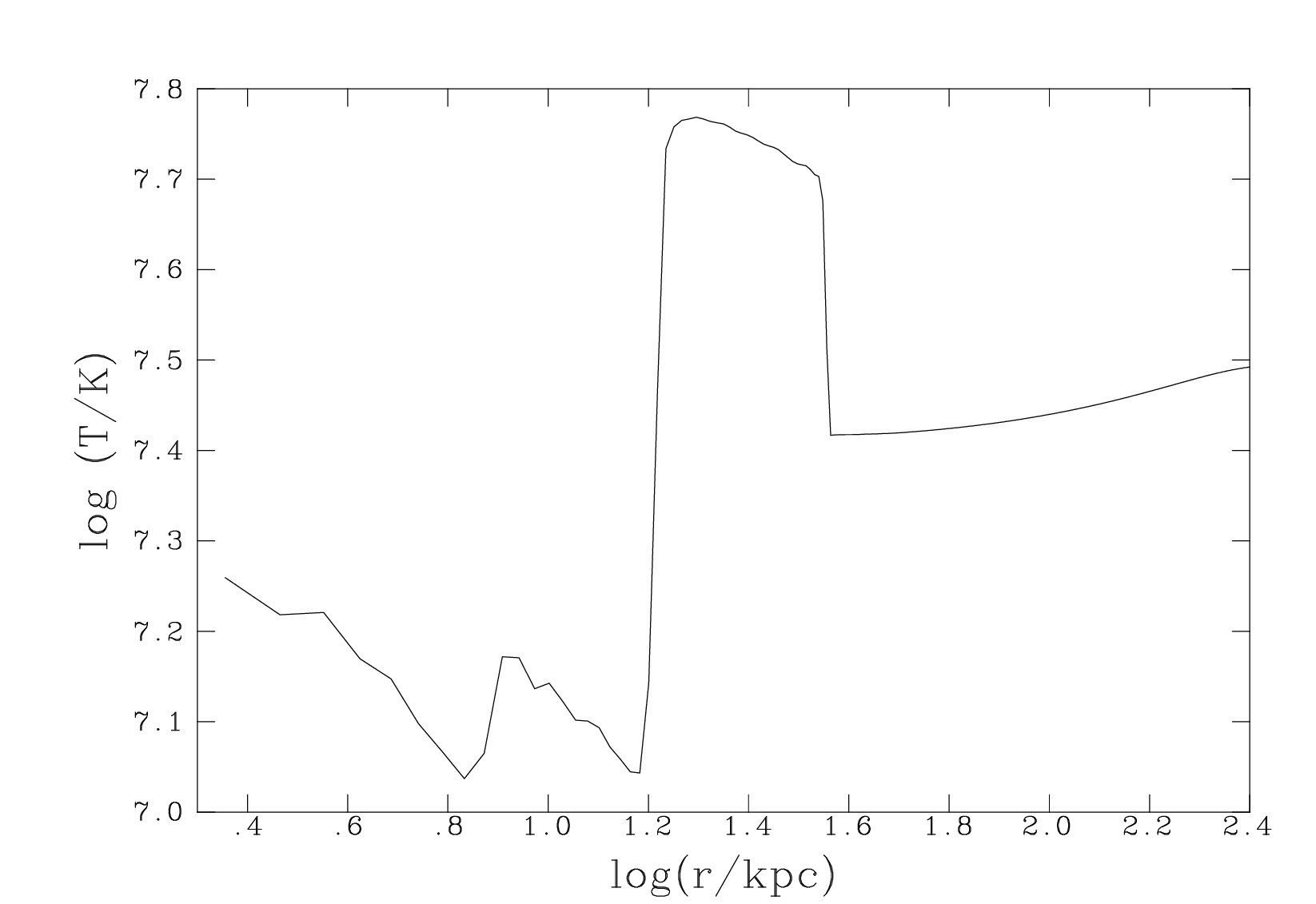}}
  \caption{On-axis surface brightness and temperature slices from the
    canonical run. These slices are taken along a line parallel to the
    direction of the galaxy motion through the galaxy centre. The centre of
    the galaxy is located at $r=0$. After 0.8 Gyr the galaxy motion is
    subsonic and a cold front is seen at $\log\left(r/\rm{kpc}\right) \sim
    1.4$, after 1.6 Gyr the galaxy motion is supersonic and a bow shock is
    seen at $\log \left(r/\rm{kpc}\right) \sim 1.6$ in addition to a cold
    front at $\log\left(r/\rm{kpc}\right) \sim 1.2$. }
  \label{fig:onaxis_slices}
\end{figure*}
Azimuthally averaged X-ray surface brightness profiles (0.3--8.0 keV) and
temperature profiles, from various times during the first and second
cluster crossings, are plotted in
Fig.~\ref{fig:surface_brightness_profiles}.  The temperature profiles are
taken directly from the simulation grid so in order for a direct comparison
to be made between these profiles and observations it would be necessary to
apply a deprojection analysis to the observed profiles.  The surface
brightness profiles consist, in general, of a central peak resulting from a
cool, dense core surrounded by the cool gas associated with the galactic
halo. This is embedded in the less dense, less bright, hotter ICM which
results in a structure reminiscent of the cold fronts seen in cluster
mergers (\citealt{vikhlinin_2001}, \citealt{markevitch_2000}).  As the
profiles in Fig.~\ref{fig:surface_brightness_profiles} are azimuthally
averaged, encompassing regions in front of and behind the galaxy, they lack
the sharp surface brightness edges seen in Chandra observations of merging
clusters.  When slices along the direction of motion, in the upstream
direction, are plotted they exhibit sharp temperature and surface
brightness edges (see Fig.~\ref{fig:onaxis_slices}). The surface brightness
slice (Fig.~\ref{fig:sb_0.8})) and the corresponding temperature slice
(Fig.~\ref{temp_0.8}) clearly indicate the presence of a cold front where a
sharp decrease in surface brightness is coincident with a transition to a
higher temperature region at $\log(r/\rm{kpc})\sim 1.4$.  The galaxy
velocity is subsonic at this time so no bow shock is present although the
gas temperature does rise approaching the discontinuity as would be
expected from adiabatic compression (\citealt{vikhlinin_2001}, \S5.3). After
1.6 Gyr the galaxy motion is supersonic and the slices 
(Fig.~\ref{fig:sb_1.6} and Fig.~\ref{temp_1.6}) are qualitatively different
to the previous subsonic case. There is still a cold front present, where a
surface brightness drop coincides with a temperature rise, but at a
slightly smaller radius ($\log(r/\rm{kpc}) \sim1.2$). In addition to the
cold front there is also a shock front where the temperature and surface
brightness both rise. The surface brightness enhancement is significantly
greater across the cold front than across the shock front which is
consistent with Chandra observations of Abell 3667 where both a cold front
and a shock front are detected but the shock front results in a smaller
surface brightness enhancement \citep{vikhlinin_2001}.  Although the size
scales and temperatures in this case are different to those encountered in
cluster mergers the gas dynamics are similar to a cluster minor merger. In
both the case of a cluster accreting a group/smaller cluster and the case
of a cluster accreting a large elliptical galaxy there is a pre-existing
halo of X-ray emitting gas, confined by a dark matter halo, at a
temperature below that of the ICM in the accreting system.

\begin{figure}
  \centering
  \includegraphics[scale=0.3, angle=-90]{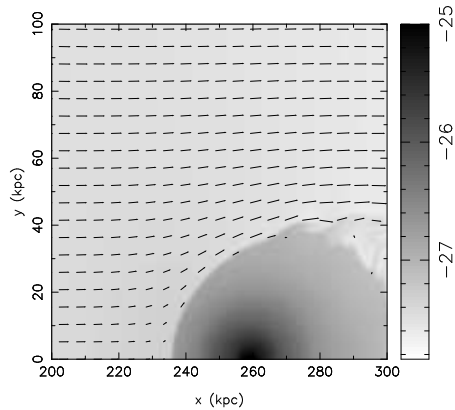}
  \caption{Density (greyscale) and velocity (vectors) around the 
    interface between the cool and hot regions. These data are taken from
    the canonical run after 0.8 Gyr.}
  \label{fig:flow_plot}
\end{figure}

\cite{vikhlinin_2001b} show that conduction, diffusion and Kelvin-Helmholtz
instabilities must be suppressed around the Abell 3667 cold front in order
for the observed sharp discontinuity to be preserved. They propose that a
magnetic field of order $10\mu\rm{G}$ would be required in a configuration
in which field lines are tangled or run parallel to the cold front.  Whilst
our simulations do not incorporate magnetic fields, conduction and
diffusion processes will be suppressed due to the absence of these terms in
the fluid equations used. Given the sharp discontinuities observed in
Chandra data the non-conductive and non-diffusive fluid equations would
seem to be appropriate.  \cite{vikhlinin_2001b} also suggest that initially
tangled magnetic fields should be stretched into a morphology where the
field lines are parallel to the front due to the velocity field of the
plasma around the interface between the cooler and hotter regions.  The
velocity field seen in our simulations (Fig.~\ref{fig:flow_plot}) clearly
demonstrates this and agrees well with the flow structure in fig.~2(b) of
\cite{vikhlinin_2001b}.

\begin{figure*}
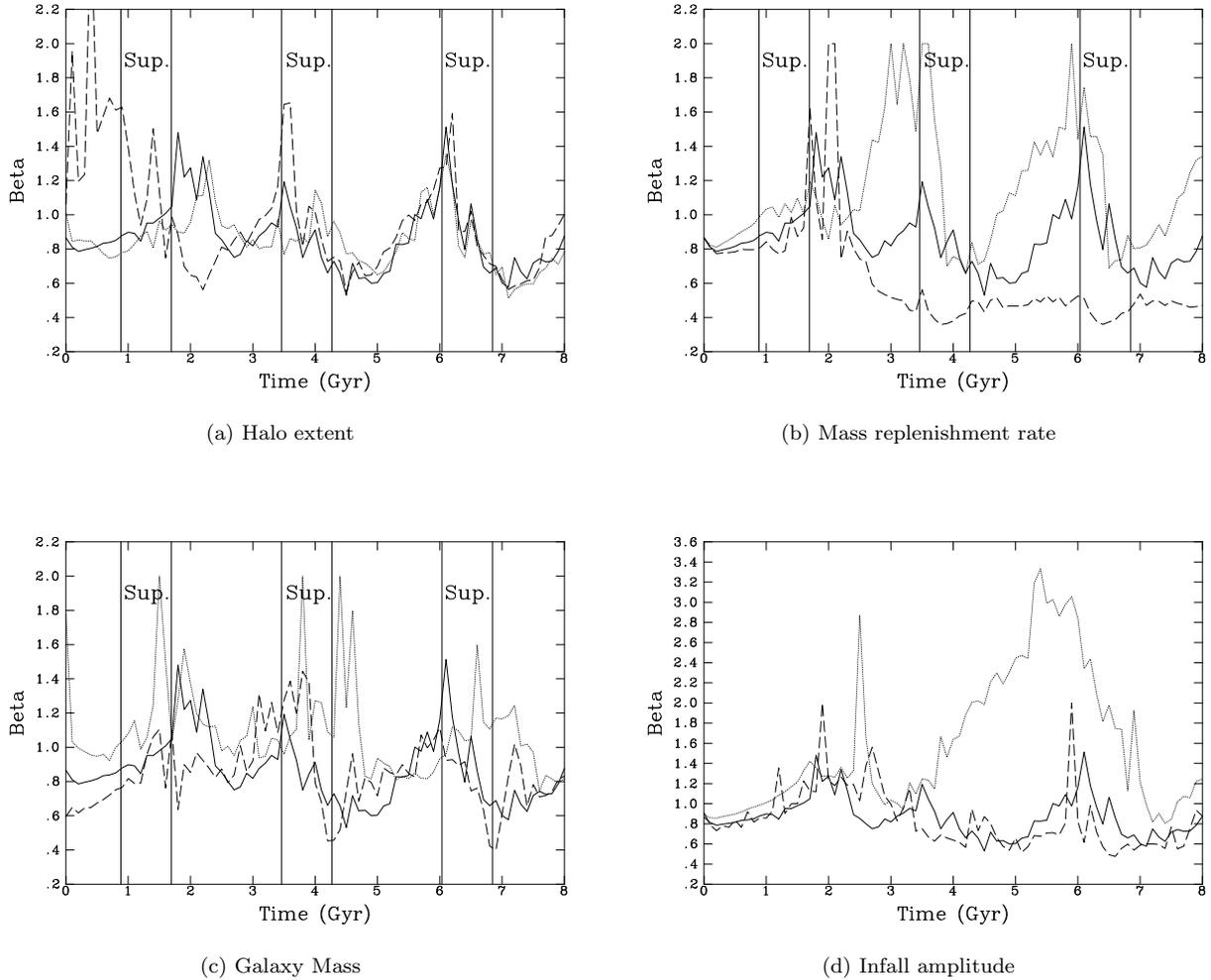

  \centering
  \subfigure[Halo extent]{
  \label{fig:beta_vs_t_halo}
  \includegraphics[scale=0.3, angle=-90]{fig13a.ps}}
  \subfigure[Mass replenishment rate]{
  \label{fig:beta_vs_t_mrep}
  \includegraphics[scale=0.3, angle=-90]{fig13b.ps}}
  \subfigure[Galaxy Mass]{
  \label{fig:beta_vs_t_mass}
  \includegraphics[scale=0.3, angle=-90]{fig13c.ps}}
  \subfigure[Infall amplitude]{
  \label{fig:beta_vs_t_infall}
  \includegraphics[scale=0.3, angle=-90]{fig13d.ps}}
  \caption{Time variation of beta values from fits to surface brightness
    profiles. Beta parameterizes the steepness of the surface brightness
    profile outside the central core region. In each case the canonical run
    is plotted as a solid line, the dashed line represents a reduction in
    the parameter and the dotted line an increase in the value of the
    parameter.}
  \label{fig:beta_vs_t}
\end{figure*}

Azimuthally averaged surface brightness profiles yield useful information
regarding the evolution of the gas halo.  The degree of central
concentration of the galaxy halo, outside the central core, may be
quantified by fitting a beta model to the profile. The surface brightness
at a projected distance $w$ from the centre of the galaxy is given by
\begin{equation}
  \label{eq:gal_beta}
  S\left(w\right) = \frac{S_0}{\left[1+\left(w/r_c\right)^2\right]^{3\beta
-\frac{1}{2}}}
\end{equation}
where $r_c$ is a core radius, $S_0$ is the central surface brightness and
the value of $\beta$ parameterizes the steepness of the profile outside the
core region.  The time variation of the beta parameter for fits to the
surface brightness distribution from all runs is plotted in
Fig.~\ref{fig:beta_vs_t}.  The model fitted to the azimuthally averaged
surface brightness profile consists of three components: a beta model to
fit the extended emission from the galaxy halo, a Gaussian to represent the
central cool core, and a second order polynomial to account for the
non-uniform cluster background.
The effect of different initial halo extents
(Fig.~\ref{fig:beta_vs_t_halo}), again shows convergence after the initial
halo has been stripped. 
During subsonic motion stripping occurs via shear instabilities at the
interface betweeen the galaxy halo and the ICM, which is less effective than
stripping in the supersonic phase. This enables replenished gas to
accumulate within the galaxy. Consequently the beta value steepens as the
regions of the halo where mass injection is effective ($r < R_H$)
become brighter.
During the
supersonic phase the beta value drops as gas is stripped and the surface
brightness profile becomes flatter. This effect is not always apparant
during the initial crossing but can be seen clearly in the third core
passage in Fig.~\ref{fig:beta_vs_t_halo}, for example.  
The effect of mass injection is seen in
Fig.~\ref{fig:beta_vs_t_mrep} where higher mass injection rates lead to
more rapidly rising beta values. This may be attributed to 
the surface brightness
profile steepening more rapidly due to the increased accumulation of gas
at radii $r<R_H$. When the infall amplitude is larger 
(Fig.~\ref{fig:beta_vs_t_infall}), the galaxy spends longer in the
outer regions of the cluster where the ram pressure is lower, so more
replenished material can accumulate and the beta value grows substantially
over an extended period of time.

\begin{figure}
  \centering
  \includegraphics[scale=0.3, angle=-90]{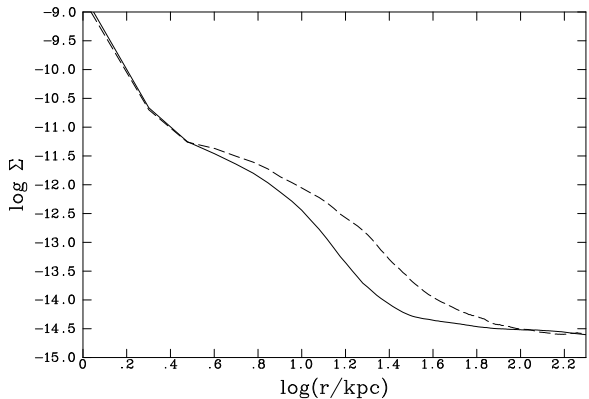}
  \caption{Comparison of the surface brightness profiles for runs with an 
    increased halo extent (dotted line) and an increased mass replenishment
    rate (solid line) after 2 Gyr in units of
    $\rm{erg}\,\rm{s}^{-1}\,\rm{cm}^{-2}\,\rm{arcmin}^{-2}.$}
  \label{fig:mrep_halo_sbcomp}
\end{figure}

The diagnostic power of surface brightness profiles, relative to integrated
X-ray luminosities, is seen when comparing the runs with an increased halo
mass and an increased mass replenishment rate. After 2 Gyr both simulated
galaxies have a similar X-ray luminosity but, as
Fig.~\ref{fig:mrep_halo_sbcomp} shows, there are marked differences in the
surface brightness profiles.
The higher mass replenishment rate, which acts only within the stellar
distribution ($\log \left(r/\rm{kpc}\right) < 1.3)$, produces a surface
brightness profile which is slightly higher in the central regions but
significantly lower in the outer regions. This information regarding the
spatial distribution of the X-ray emitting gas is lost when considering
only integrated X-ray luminosities.

\subsection{Observable wake properties}
\label{subsec:wake}
\begin{figure*}
  \centering
  \subfigure[First crossing, t=1.3 Gyr]{
  \label{fig:sbimage_canonical_first}
  \includegraphics[scale=0.18, angle=-90]{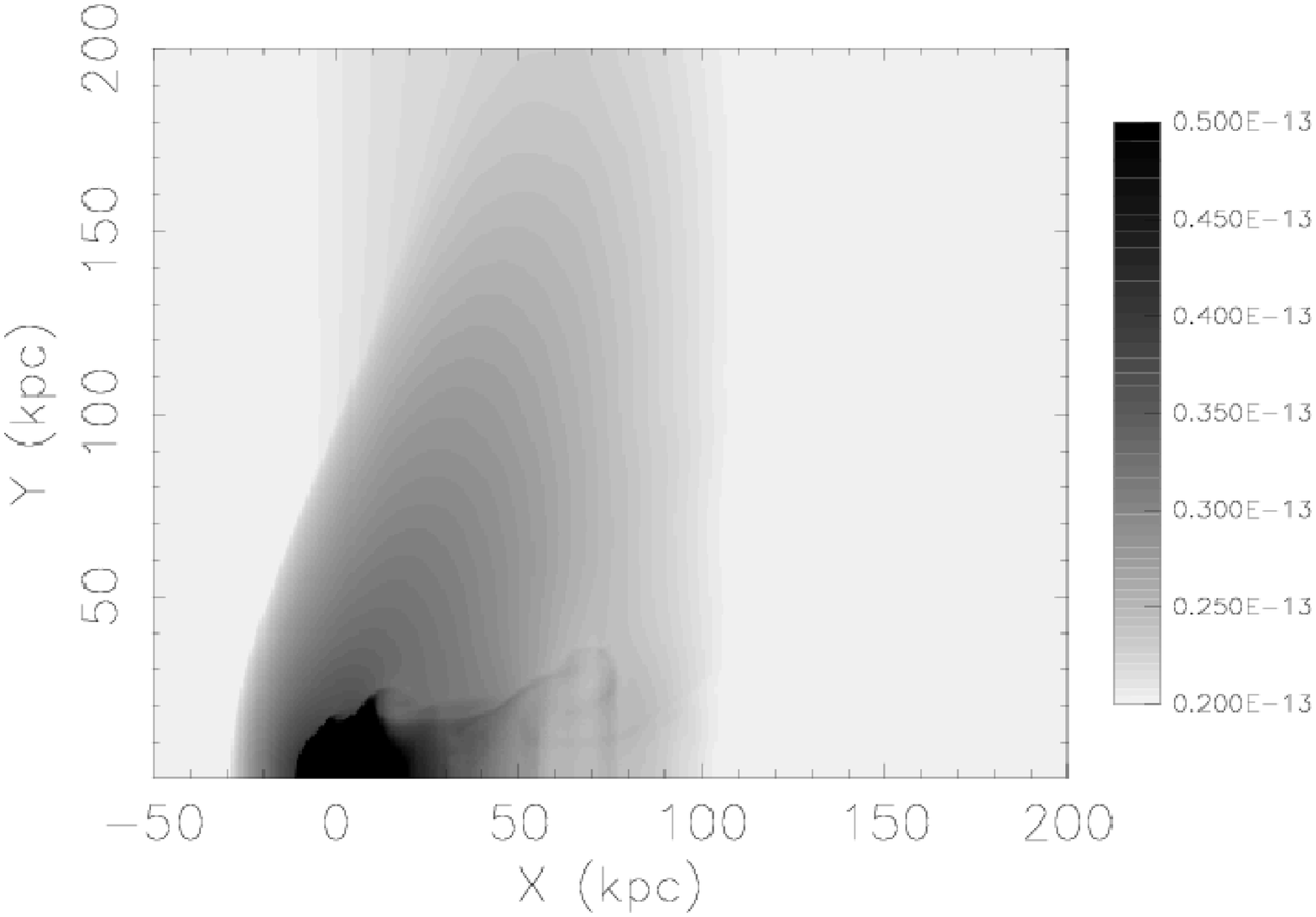}}
  \subfigure[Second crossing, t=3.9 Gyr]{
  \includegraphics[scale=0.18, angle=-90]{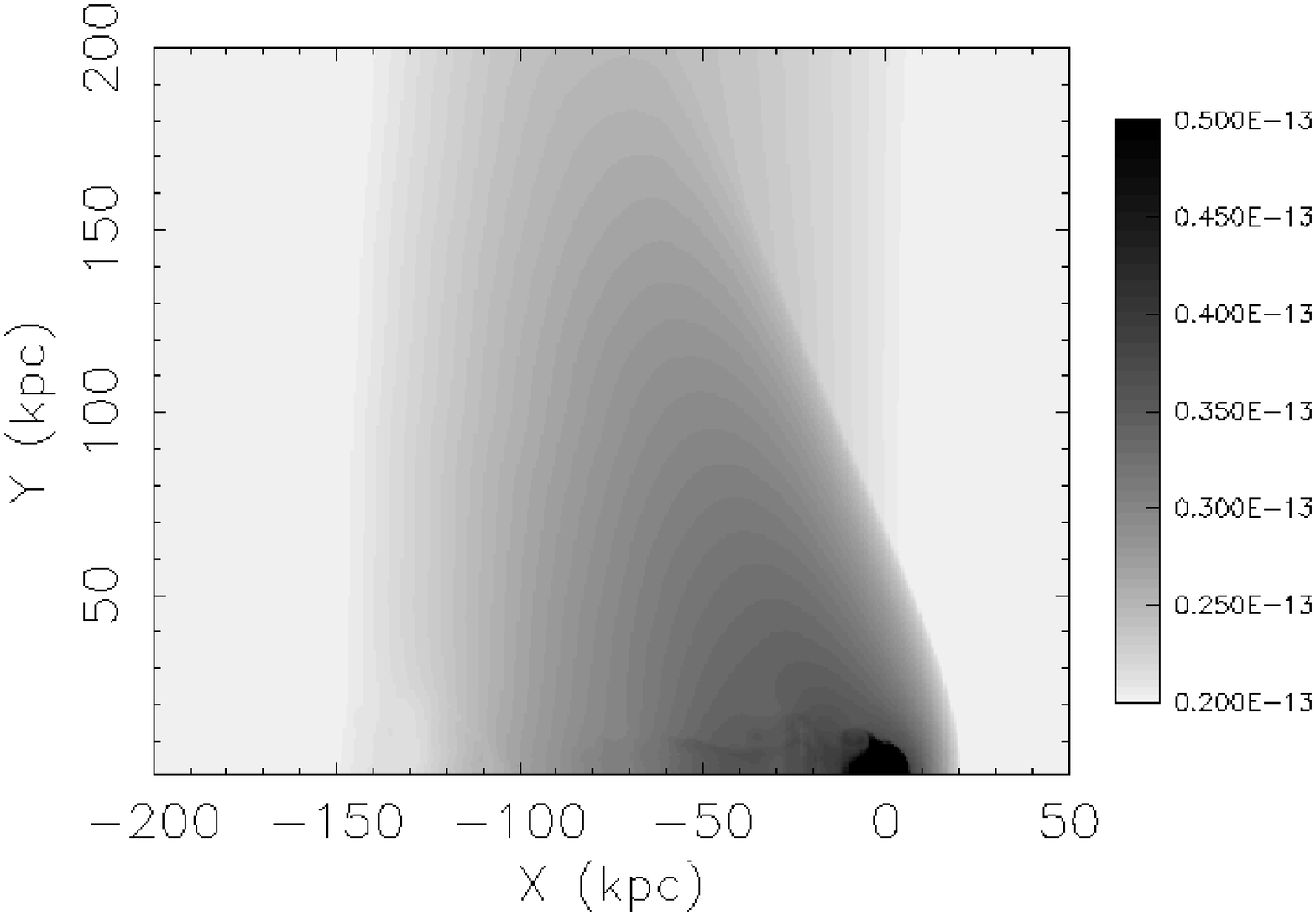}}
  \subfigure[Third crossing, t=6.4 Gyr]{
  \includegraphics[scale=0.18, angle=-90]{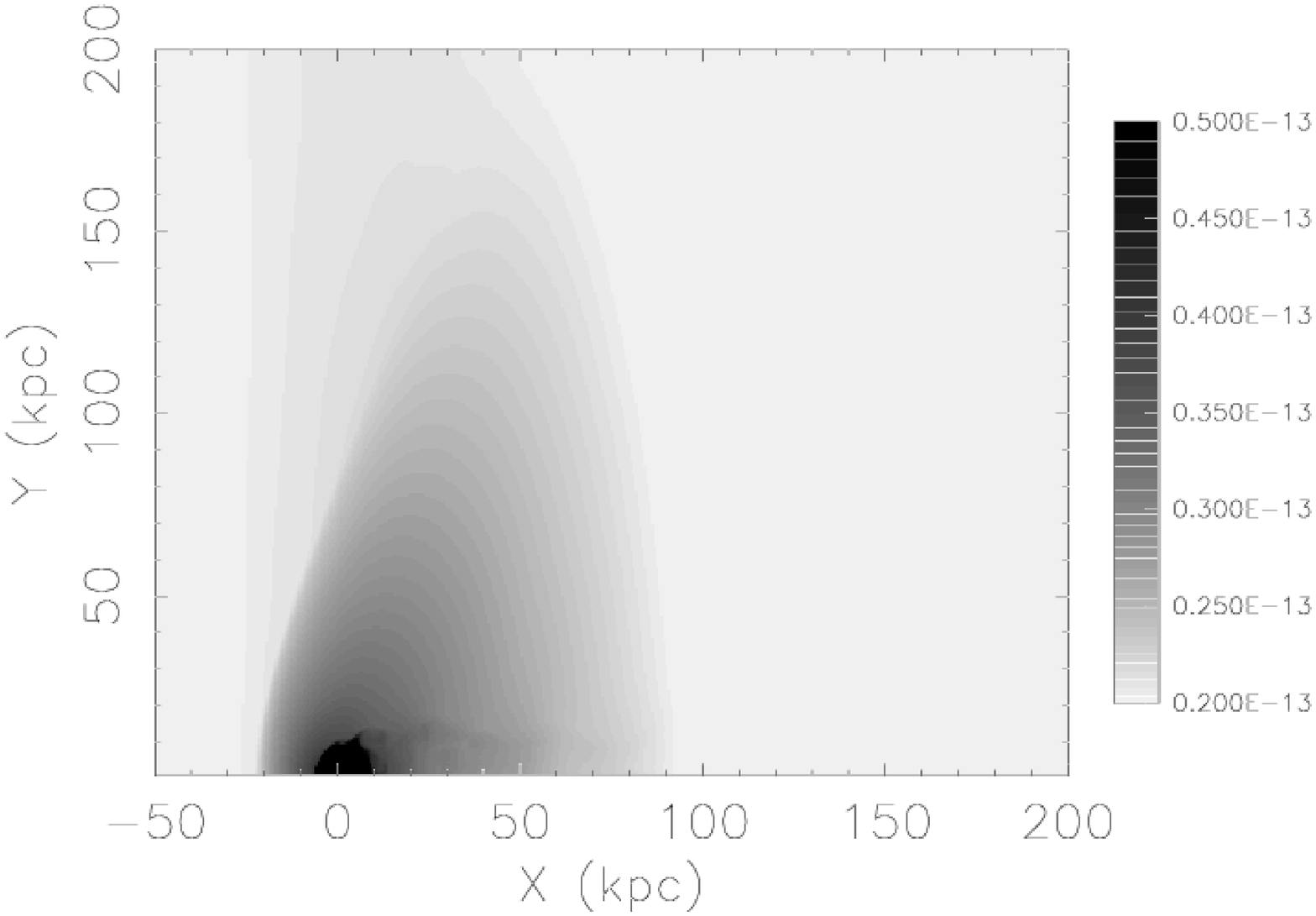}}
  \subfigure[First crossing, t=1.3 Gyr]{
  \label{fig:sbimage_larger_halo_first}
  \includegraphics[scale=0.18, angle=-90]{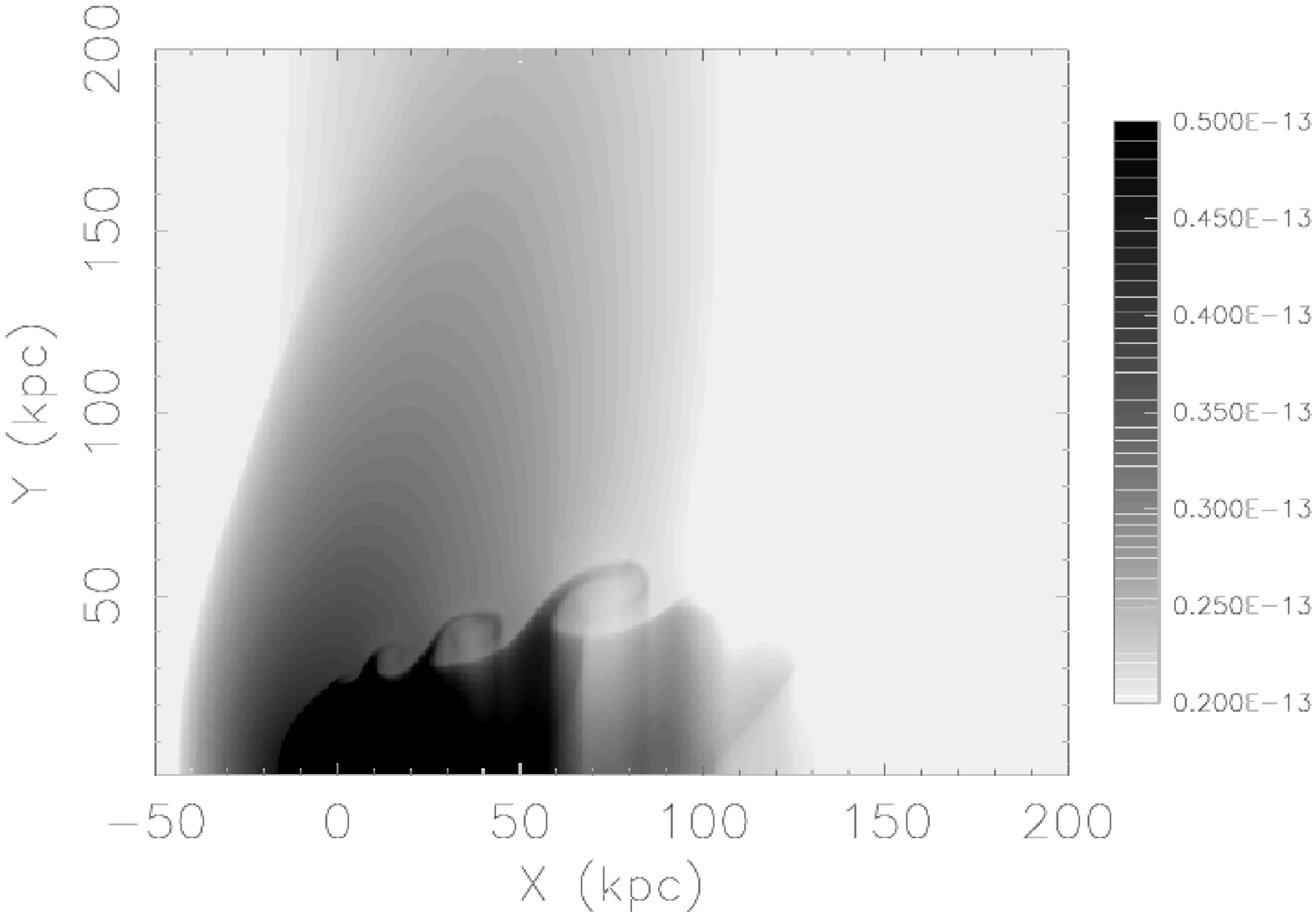}}
  \subfigure[Second crossing, t=3.9 Gyr]{
  \includegraphics[scale=0.18, angle=-90]{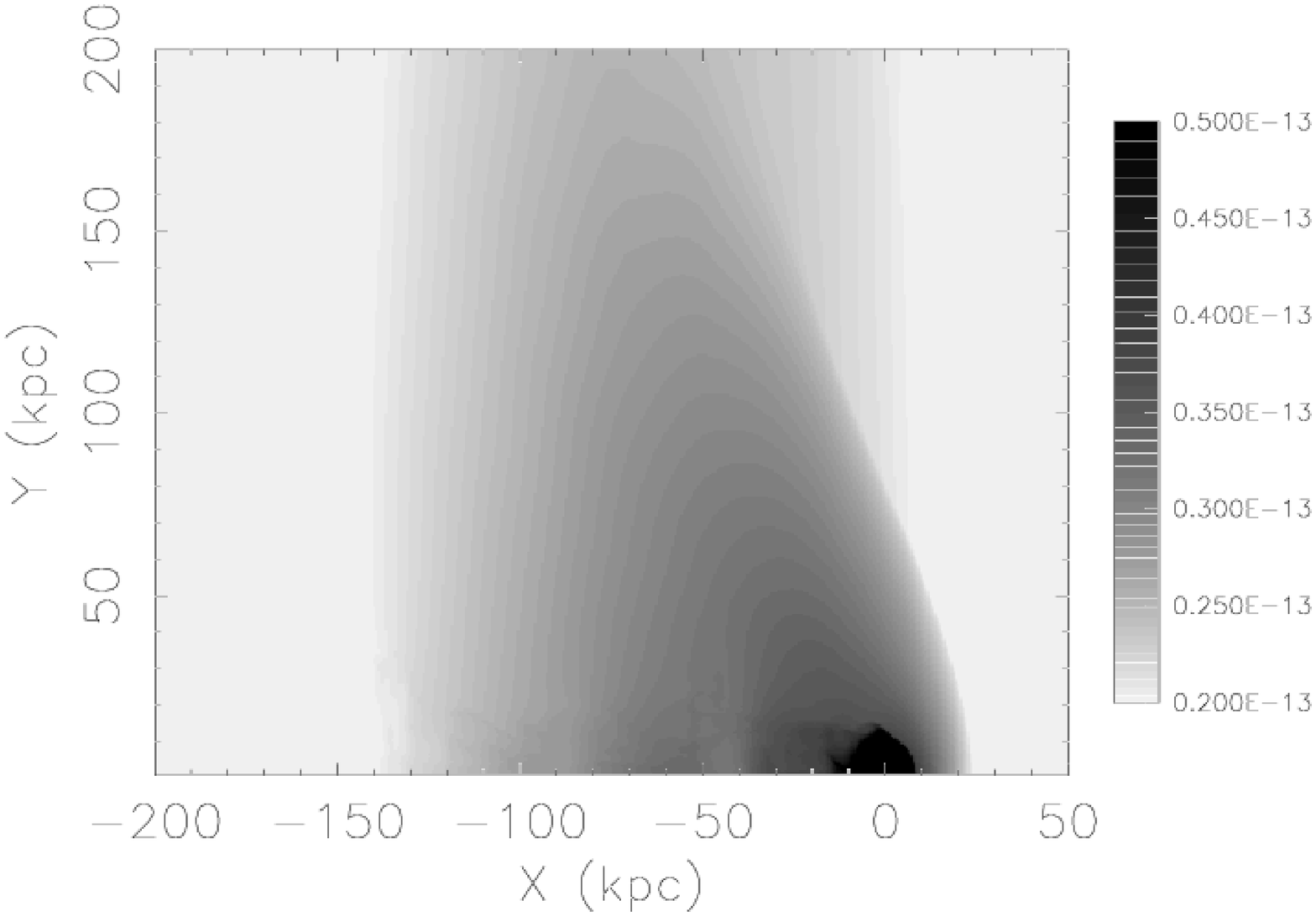}}
  \subfigure[Third crossing, t=6.4 Gyr]{
  \includegraphics[scale=0.18, angle=-90]{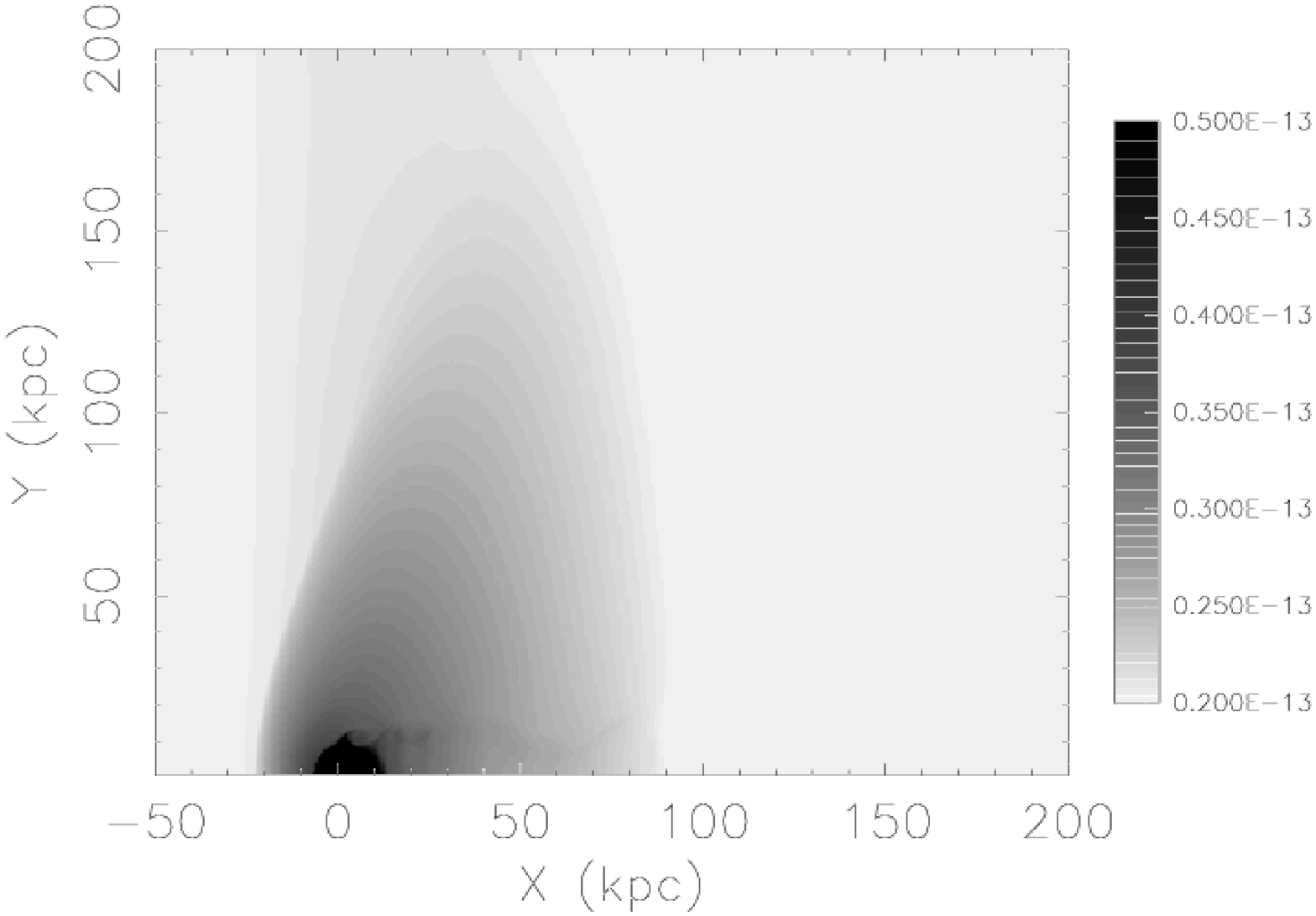}}
  \caption{Simulated X-ray flux maps from the canonical run (top) and the
    larger gas halo run (bottom). 
  All images are plotted using the same logarithmic greyscale in units of $\rm{erg}\,\rm{s}^{-1}\,\rm{cm}^{-2}\,\rm{arcmin}^{-2}.$}
  \label{fig:sbimages}
\end{figure*}

In addition to the X-ray emission from the body of the galaxy there is also
emission from stripped material downstream of the galaxy which may form an
observable wake. The top row of Fig.~\ref{fig:sbimages} shows X-ray surface
brightness maps
from the canonical simulation during the first, second and third crossings
of the cluster at the time when the galaxy velocity is at a maximum. The
highest surface brightness feature in all cases is the emission from the
body of the galaxy and the next brightest is 
the tail of stripped material. The surface
brightness of the tail is much higher during the first crossing than in
subsequent crossings indicating a bias towards observing wakes around
galaxies on their initial infall into a cluster.
The bottom row of Fig.~\ref{fig:sbimages} shows surface brightness maps
from the run with a larger initial gas halo. The wake formed during the
first cluster crossing is brighter and wider than that from the canonical
run suggesting that the initial gas content of the galaxy has a major
influence on whether wakes will be observable.
\begin{table}
  \caption{Wake luminosities during the first cluster crossing when the
    galaxy velocity is maximal (i.e.\ at core passage).}
  \centering
  \begin{tabular}{lc}
\hline
Run                              &  $\log\left(L_X\right)$  \\
\hline 
Canonical                        &  40.9 \\
Larger halo                      &  41.3 \\
Smaller halo                     &  40.4 \\
Larger mass replenishment rate   &  40.9 \\
Smaller mass replenishment rate  &  40.9 \\
Larger galaxy                    &  40.9 \\
Smaller galaxy                   &  40.6 \\
Larger infall amplitude          &  39.9 \\
Smaller infall amplitude         &  40.4 \\
\hline
  \end{tabular}
  \label{tab:tail_lx}
\end{table}
X-ray luminosities, in a 0.3--8.0 keV band, for the first cluster crossing
are tabulated in Table~\ref{tab:tail_lx}.  These luminosities have been
calculated as enhancements above the background level which would be
present at the same distance from the cluster centre but away from the
galaxy location.  The brightness of the wake is most strongly affected by
the pre-existing halo size; larger initial haloes result in more luminous
wakes. The mass replenishment rate does not have a noticable influence on
wake luminosity during the first infall. Its effect only becomes apparant
at later times when it affects the X-ray luminosity from the body of the
galaxy. 

\begin{figure*}
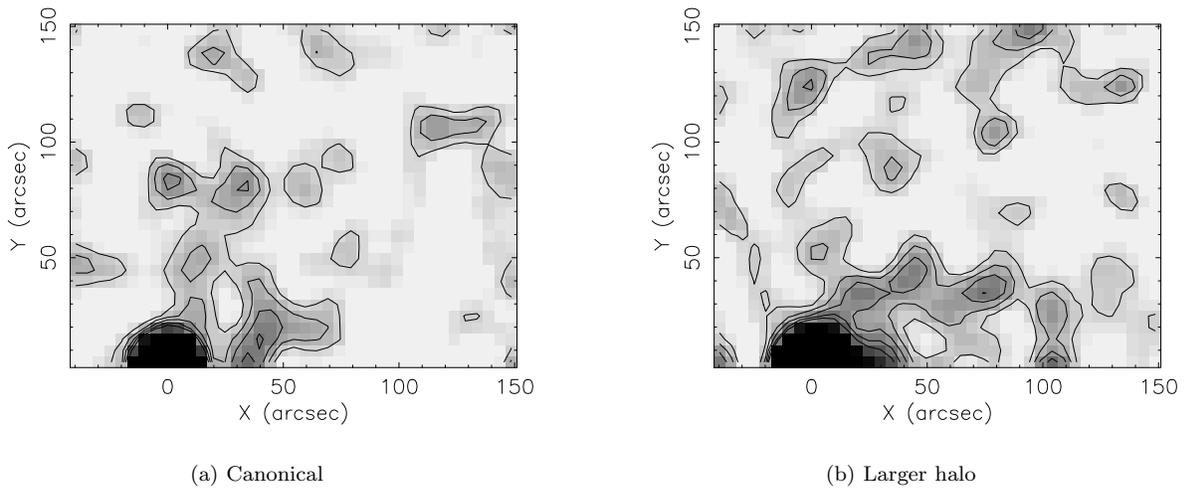

  \centering
  \subfigure[Canonical]{
  \includegraphics[scale=0.3,angle=-90]{fig16a.ps}}
  \subfigure[Larger halo]{
  \includegraphics[scale=0.3,angle=-90]{fig16b.ps}}
  \caption{Significance levels in simulated X-ray images 
    derived using the flux maps show in
    Fig.~\ref{fig:sbimage_canonical_first} and
    Fig.~\ref{fig:sbimage_larger_halo_first}.  In both cases the galaxy is
    at its first core passage, after 1.3 Gyr of simulated infall.  Contours
    represent significance and are at levels of 1,2,3,4,5 $\sigma$.}
  \label{fig:simim_1013}
\end{figure*}
The X-ray flux maps presented in Fig.~\ref{fig:sbimages} are valuable for
determining which features result in the most observable X-ray emission.
However, they should be treated with caution as they do not include the
effects of detector sensitivity, spatial resolution and background levels.
Flux maps, such as those in Fig.~\ref{fig:sbimages}, may be converted to
simulated X-ray images as shown in Fig.~\ref{fig:simim_1013}. These
simulated images represent 50~ksec Chandra observations of the canonical
galaxy at first core passage (based on the flux map in
Fig.~\ref{fig:sbimage_canonical_first}) and the galaxy with a larger
initial halo, also at first core passage (based on the flux map in
Fig.~\ref{fig:sbimage_larger_halo_first}).  In both cases the cluster is
placed at a distance of 270~Mpc. An energy band of 0.8--2.0 keV is used to
increase the contrast between the cooler galaxy emission and the hotter
cluster background.  The spatial bins are 5 arcsec square and 
$1\rm{kpc}=0.76~\rm{arcsec}$ at the assumed distance of 270~Mpc.  The
effects of detector background and Poisson noise have been added, a cluster
background level is then subtracted and the resulting image smoothed with a
Gaussian filter of 3 pixels FWHM. Even at this relatively large distance a
tail of material is visible in both cases, most prominently in the larger
halo case, although it is affected by noise in the outer regions. These
simulated images confirm the observability of the structures seen in the
flux maps: the galaxy halo is the most observable feature, the stripped
wake is the next brightest structure (with observability declining with
increasing distance from the galaxy) and the bow shock is the least
observable feature, but may still be detectable.

\begin{figure*}
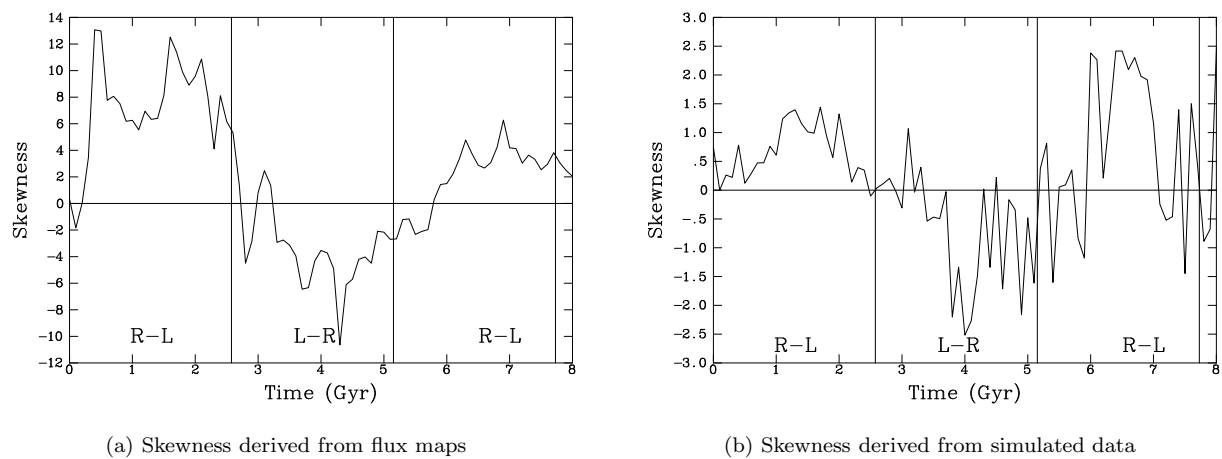

  \centering
  \subfigure[Skewness derived from flux maps]{
  \label{fig:skewness_fluxmap}
  \includegraphics[scale=0.3,angle=-90]{fig17a.ps}}
  \subfigure[Skewness derived from simulated data]{
  \label{fig:skewness_simdata}
  \includegraphics[scale=0.3,angle=-90]{fig17b.ps}}
  \caption{Time variation of skewness parameter based on the flux maps from
    the canonical case. The vertical lines mark regions where the galaxy
    motion is from right to left (R-L) or from left to right (L-R). When
    the skew is positive the emission is distorted to the right and when
    the skew is negative the distortion is to the left.}
  \label{fig:skewness}
\end{figure*}

A simple statistical measure of the degree of asymmetry present in the
galaxy's X-ray halo can be formed using skewness
\begin{eqnarray}
  \label{eq:skew}
  \gamma = \frac{1}{F_{\rm{T}}\sigma^3} \sum_{i,j} F_{i,j} \left( x_{i,j}
    -\bar{x} \right)^3 \\
  \sigma^2 = \frac{1}{F_{\rm{T}}} \sum_{i,j} F_{i,j} \left( x_{i,j}
    -\bar{x} \right)^2
\end{eqnarray}
where $i$ and $j$ are indices representing cells along the $x$ and $y$ axes
respectively, $F_{i,j}$ is the flux from an individual cell, $F_{\rm{T}}$
is the total flux from the image, $x_{i,j}$ is the location of the cell
along the $x$-axis and $\bar{x}$ is the $x$ location of the galaxy centre
along the $x$-axis. The time variation of the skewness parameter $\gamma$,
for the larger halo simulation is plotted in
Fig.~\ref{fig:skewness_fluxmap}. This has been calculated using the broad
band flux maps. Skewness derived from the corresponding simulated images is
plotted in Fig.~\ref{fig:skewness_simdata} for the same run.
Fig.~\ref{fig:skewness_fluxmap} represents an idealised case with no noise
whereas Fig.~\ref{fig:skewness_simdata} shows the effects of noise present
in real data.  When the value of skewness is positive the emission is
distorted to the right and the skewness is negative when the emission is
distorted to the left.  Skewness is, in general, a good measure of the
direction of galaxy motion although when the galaxy speed is low, in the
outer regions of the cluster, there are periods when the skewness does not
indicate the correct direction of motion. Given a sample of galaxies from a
cluster the skewness parameter will serve as a statistical indicator of the
direction of motion with respect to the ICM even after the first infall
phase. A sample drawn from the central regions of the cluster will
preferentially select cases where the skewness measure is a reliable
direction indicator. In the above examples skewness has been calculated in
a circular region of radius 120 arcsec ($\sim 90 \rm{kpc}$) centred on the
galaxy to approximately encompass the significant emission from any wake
formed (c.f.\ Fig.~\ref{fig:sbimage_larger_halo_first}). When applying
skewness to real data the reliability of the measure may be enhanced by
tuning the size of the region used relative to the extent of galactic
haloes seen in the observation.

\subsection{Shock properties}
\label{subsec:shock}
\begin{figure*}
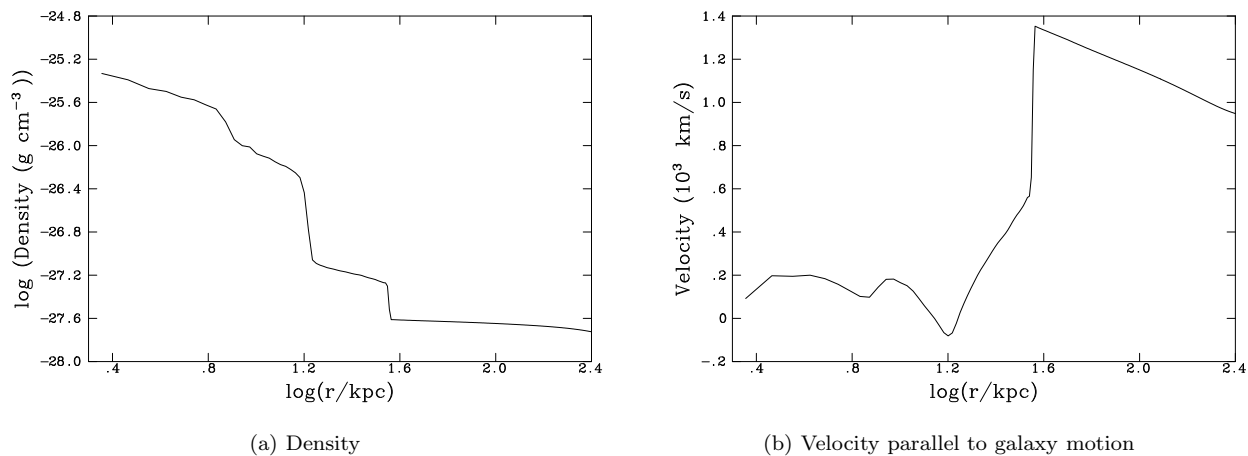

  \centering
  \subfigure[Density]{
  \label{fig:shock_density}
  \includegraphics[scale=0.3,angle=-90]{fig18a.ps}}
  \subfigure[Velocity parallel to galaxy motion]{
  \label{fig:shock_vex}
  \includegraphics[scale=0.3,angle=-90]{fig18b.ps}}
  \caption{Density and velocity jumps across the shock 
    after 1.6 Gyr. The both quantities are measured upstream, parallel to
    the direction of galaxy motion along a line through the centre of the
    galaxy. The galaxy is centred at $r=0$.}
  \label{fig:shock}
\end{figure*}

The density and temperature jumps across a shock, such as that which
precedes the galaxy during supersonic motion, can be derived from the
Rankine-Hugoniot relations \citep{landau_1959},
\begin{equation}
  \label{eq:dens_jump}
  \frac{\rho_2}{\rho_1} = \frac{\left(1+\gamma\right) M^2_1} 
  {2+\left(\gamma - 1\right)M^2_1}
\end{equation}
\begin{equation}
  \label{eq:temp_jump}
  \frac{T_2}{T_1} = \frac{2\gamma M^2_1 -\gamma+1}{\gamma+1}
  \frac{\rho_1}{\rho_2}
\end{equation}
where $\gamma=5/3$ for the monatomic gas considered here. The density ratio
across the shock front, present after 1.6 Gyr, is measured to be 2.0 (see
Fig.~\ref{fig:shock_density}) which indicates that the Mach number of the
incident gas is $\sim1.7$. The local gas temperature in the region
immediately upstream of the shock is $\sim2.6\times10^7$ K (see
Fig.~\ref{temp_1.6}) indicating that the gas velocity
immediately upstream of the shock is $\sim1330 \, \rm{km\,s^{-1}}$, consistent
with the velocity profile in Fig.~\ref{fig:shock_vex}. The predicted
temperature ratio across the shock is 1.8 consistent with that observed in
Fig.~\ref{temp_1.6}. 
The velocity and Mach number derived above are local values which pertain
to the region immediately upstream of the shock. The galaxy is a
gravitating body and causes gas upstream to accelerate towards the shock
location hence the velocity of the gas calculated above is higher than the
global velocity of the galaxy with respect to the host cluster.
Additionally there is a decrease in the local gas temperature during the
acceleration towards the shock which also causes the Mach number to
increase above the global value. In the above example the velocity of the
galaxy with respect to the host cluster is 943 $\rm{km\,s^{-1}}$ but the
value derived from the density jump is $1330 \, \rm{km\,s^{-1}}$.

\section{Stripping in a larger cluster}
\label{sec:hotcl}

The parameter study of a number of scenarios in the same modest cluster
yields information about how a homogeneous sample of wakes may be biased.
However, it is also instructive to investigate the effect of placing the
model galaxy in a more substantial cluster where stripping is expected to
be more effective.  Thus we performed a simulation with the canonical
galaxy placed in a cluster with the cluster temperature scaled up by a
factor of 2 to 5.4 keV. Several parameters had to be scaled accordingly, so
that the cluster remains realistic. The scaled values are shown in
Table~\ref{tab:hotter_clust}. We scale the cluster velocity dispersion
$\sigma_c$, the cluster core radius $r_c$ and the central ICM density
$n_{\rm{ICM}}$ as follows.  
\begin{equation}
    \sigma_c  \propto T^{1/2}
\end{equation}
to be consistent with eqn.~\ref{eq:sigma}, and 
\begin{equation}
  r_c \propto T^{1/2}       
\end{equation}
assuming that the core radius is a fixed fraction of the virial radius and
that the virial radius scales with temperature in a self similar fashion.
The scaling of the central ICM density was based on the X-ray emission
properties. For a beta model surface brightness distribution with
$\beta=2/3$ the central emissivity $S_0$ scales as
\begin{equation}
  S_0 \propto j_0 r_c
\end{equation}
where the central emissivity $j_0$ is assumed to scale as 
\begin{equation}
   j_0 \propto n^{2}_{\rm{ICM}}
\end{equation}
hence
\begin{equation}
  \label{S_vs_nr}
  S_0 \propto n^{2}_{\rm{ICM}} r_c.
\end{equation}
From eqn.~\ref{eq:lx_sb}
\begin{equation}
\label{L_vs_Sr}
L_X \propto S_0 r_c^2.
\end{equation}
Combining eqns.~\ref{S_vs_nr} and \ref{L_vs_Sr} gives 
\begin{equation}
  n^{2}_{\rm{ICM}} \propto L_{\rm{X}} r_c^{-3}.
\end{equation}
It is assumed that $\rm{L_{X}}$ scales as $\rm{L_X} \propto T^3$ in
accordance with observationally determined relations \citep{white_1997} 
rather than $\rm{L_X} \propto T^2$ which would be expected from self
similarity.
\begin{table}
  \centering
  \caption{Parameters of the hotter cluster.}
  \begin{tabular}[lc]{lc}
\hline
Parameter & Value \\
\hline
Temperature            & $5.4\,\rm{keV}$ \\
Velocity dispersion    & $928\,\rm{km\,s^{-1}}$ \\
Core radius            & $566\,\rm{kpc}$ \\
Central ICM density    & $1.75\times 10^{-3}\,\rm{cm}^{-3}$ \\
\hline
  \end{tabular}
  \label{tab:hotter_clust}
\end{table}

\begin{figure}
  \centering
  \includegraphics[scale=0.3, angle=-90]{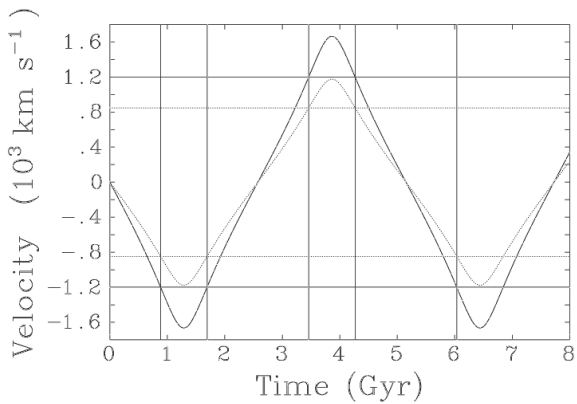}
  \caption{Velocity of the galaxy in the hotter cluster (solid line) and
    the 2.7 keV cluster (dotted line). Horizontal lines mark the cluster
    sound speed (solid line is hotter cluster, dotted line is 2.7 keV
    cluster). Solid vertical lines mark the transition from the subsonic to
    supersonic regime which is the same in both cases.}
  \label{fig:hotcl_dyn}
\end{figure}

The velocity of the galaxy in the hotter cluster is plotted as the solid
line in Fig.~\ref{fig:hotcl_dyn}, and for comparision the velocity
variation for the galaxy in the 2.7 keV cluster is shown by the dotted
line. The cluster sound speed is shown by the solid and dotted horizontal
lines for the hotter cluster and cooler cluster respectively. The cluster
crossing time is the same in both cases as galaxy velocity and cluster size
both scale as $T^{1/2}$.

\begin{figure}
  \centering
  \includegraphics[scale=0.3, angle=-90]{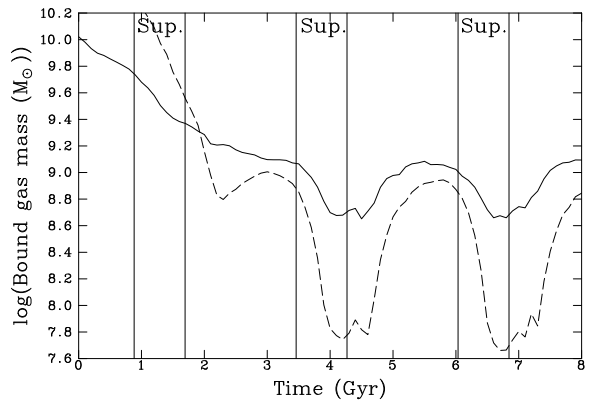}
  \caption{Mass of gas gravitationally bound to the galaxy in the canonical
    (solid line) and hotter cluster (dashed line) runs.}
  \label{fig:boundgas_clust}
\end{figure}

The mass of gas gravitationally bound to the galaxy (shown in
Fig.~\ref{fig:boundgas_clust}) is significantly less in the hotter cluster
after the first 2 Gyr.  The bound gas content is initially higher as the
galactic gas halo density must increase to counteract the increased ICM
pressure confinement.  However, this extra gas is quickly stripped during
the first crossing.  The ram pressure is higher in a hotter cluster due to
the higher ICM density and also due to the higher velocities reached.  As a
consequence it would be expected that galaxies residing in such an
environment would retain less gas than those galaxies in cooler clusters
where stripping is less effective. 

It is expected that wakes will be more difficult to observe in hotter
clusters for two reasons. Firstly, the gas content of the cluster galaxies
should be rapidly depleted by stripping, and secondly the background
emission from the ICM will be brighter than in a cooler system. 
\begin{figure*}
  \centering 
    \subfigure[Simulated X-ray flux map similar to those in
    Fig.~\ref{fig:sbimages}. Note the difference in the greyscale levels used
    here compared to those used in Fig.~\ref{fig:sbimages}]{
    \includegraphics[scale=0.3,angle=-90]{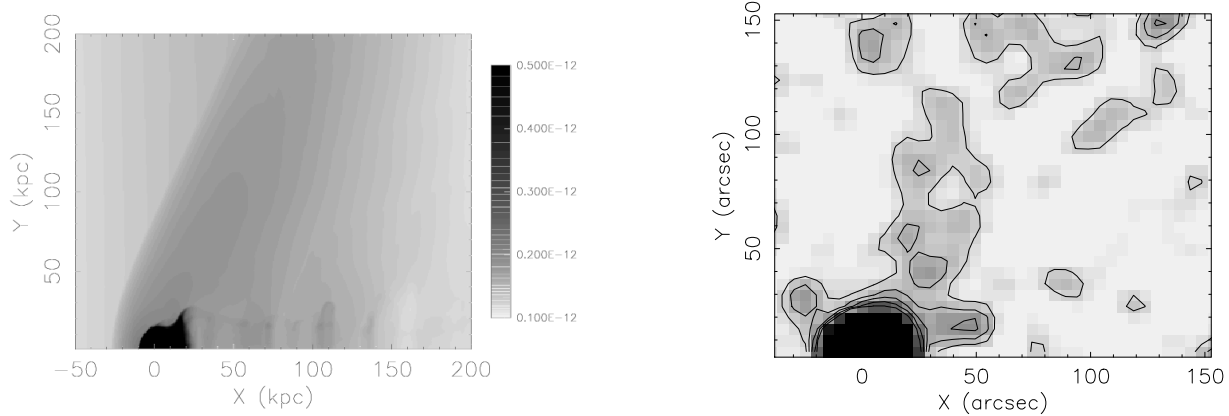}}
    \subfigure[Significance levels in a simulated X-ray image, similar those
    in Fig.~\ref{fig:simim_1013}. Contours represent significance and are at
    levels of 1,2,3,4,5 $\sigma$.]{
    \label{fig:hotcl_sig}
    \includegraphics[scale=0.3,angle=-90]{fig21b.ps}}
    \caption{Simulated X-ray flux map and significance map for the hotter
    cluster run after 1.3 Gyr.}
  \label{fig:hotcl_obs}
\end{figure*}
In order to assess the observability of wakes in the hotter cluster we have
again derived simulated X-ray flux maps and simulated images. These are
presented in Fig.~\ref{fig:hotcl_obs}. The flux map was generated in the
same way as those in Fig.~\ref{fig:sbimages} and the significance plot was
generated in the same way as Fig.~\ref{fig:simim_1013} to enable a direct
comparison between the results for the hotter and canonical clusters. The
gas halo associated with the galaxy produces a significant amount of
emission, as seen in Fig.~\ref{fig:hotcl_sig}, as the high initial gas
content has not yet been stripped at this stage. There is a wake of
emission from stripped material but it is shorter and narrower than the
previous wakes in Fig.~\ref{fig:simim_1013} indicating that wakes in hotter
clusters will be less observable than in cooler systems. The bow shock may
also be detectable.

\section{Summary and discussion}
\label{section:conclusions}

We present results from a parameter study involving hydrodynamic
simulations of a large elliptical galaxy falling radially into an
intermediate temperature cluster. The resulting stripping causes variations
in the gas mass associated with the galaxy and in the resulting X-ray
luminosity. Variations in model parameters (particularly mass replenishment
rate) are found to induce larger variations in $\rm{L_X}$ ($\sim2$ dex)
than the effects of observing a given model at different points during the
galaxy's dynamic cycle.

The luminosity of the X-ray wake during the first cluster passage is found
to be most strongly influenced by the extent of the pre-existing gas halo
and during subsequent passages through the cluster there is not a
bright X-ray wake. This indicates that samples of wakes in clusters will be
biased towards large elliptical galaxies, with substantial haloes, which
are on their first passage through the cluster.  Wakes surrounding galaxies
residing in hotter, more massive clusters are, as expected, less easily
detected.

The infall of a cooler gas halo into a cluster results in structures
reminiscent of the cold fronts seen in cluster mergers. The cold front is
found to be a more prominent surface brightness feature than the shock
front, present during supersonic motion, in agreement with recent Chandra
results of cluster mergers. The density and temperature ratios across the
shock front are found to be consistent with values from analytic theory but
the velocity derived in this way is a local value which differs from the
global velocity of the galaxy with respect to the cluster potential as the
galaxy is a gravitating body which accelerates the local ICM.  Surface
brightness enhancements and temperature variations arising from from the
galaxy motion should be detectable with present instrumentation such as
Chandra and XMM.

\section*{Acknowledgements}

DMA acknowledges funding from a Postgraduate Teaching Assistantship from
the School of Physics \& Astronomy, while IRS and IS acknowledge funding
from PPARC. We would like to thank an anonymous referee for helpful
comments. 

\bibliographystyle{mn2e}
\bibliography{bib}

\end{document}